\documentclass[journal]{IEEEtran}

\usepackage{cite}
\usepackage{amsmath,amssymb}
\usepackage{graphicx}

\usepackage{booktabs}
\usepackage{tabularx}
\usepackage{array}

\usepackage[caption=false,font=footnotesize]{subfig}
\usepackage{stfloats}
\usepackage{url}

\graphicspath{{figures/}}
\begin{document}

\title{A Reconfigurable Pipelined-SAR ADC with Embedded Compression for Temporal Compressed-Sensing Ultrasound Imaging}

\author{Reza Pakdaman Zangabad\IEEEauthorrefmark{1},~\IEEEmembership{Member,~IEEE},
Xitie Zhang\IEEEauthorrefmark{1},~\IEEEmembership{Student Member,~IEEE},\\
Levent Degertekin,~\IEEEmembership{Fellow,~IEEE},
and Shaolan Li,~\IEEEmembership{Senior Member,~IEEE}
\thanks{This work was supported by the National Institutes of Health (NIH), under Grant 1R21EB034991.}
\thanks{\IEEEauthorrefmark{1}Reza Pakdaman Zangabad and Xitie Zhang contributed equally to this work.}
\thanks{R. Pakdaman Zangabad is with the Mechanical Engineering Department, Georgia Institute of Technology, Atlanta, GA 30332 USA (e-mail: rpz3@gatech.edu).}
\thanks{X. Zhang is with the Electrical and Computer Engineering Departments, Georgia Institute of Technology, Atlanta, GA 30332 USA (e-mail: xitiezhang@gatech.edu).}
\thanks{S. Li is with the Electrical and Computer Engineering Departments, Georgia Institute of Technology, Atlanta, GA 30332 USA (e-mail: shaolan.li@ece.gatech.edu).}
\thanks{L. Degertekin is with the Mechanical and Electrical and Computer Engineering Departments, Georgia Institute of Technology, Atlanta, GA 30332 USA (e-mail: levent.degertekin@me.gatech.edu).}
\thanks{Corresponding author: Reza Pakdaman Zangabad (e-mail: rpz3@gatech.edu).}
}

\markboth{arXiv,~Vol.~XX, No.~XX, 2026}%
{Pakdaman Zangabad \MakeLowercase{\textit{et al.}}: Compressed Sensing Medical Ultrasound Imaging Using Piplined SAR ADC}

\maketitle

\begin{abstract}
Compact ultrasound imaging systems are increasingly constrained by
receiver-side sampling, conversion, memory, and data-transfer
requirements. This work presents a compressed-sensing pipelined
successive-approximation-register analog-to-digital converter (CS-SAR
ADC) for acquisition-side temporal compression of pre-beamformed medical
ultrasound radio-frequency (RF) data. Pseudo-random polarity modulation
and charge-domain accumulation are embedded in the SAR sampling network
so that multiple consecutive RF samples are encoded into one measurement
before quantization, supporting temporal compression ratios of
$N_{cT}=1$, 2, and 4. The compressed outputs are recovered off chip using
a probe-specific pulse-dictionary RF model and then processed with
conventional ultrasound beamforming. A 65-nm CMOS prototype was measured
with a 1.2-V supply and 50-MHz master clock. In the non-compressed mode,
it operates at 10 MS/s, consumes 964.49~$\mu$W, and achieves 44.12-dB
SNDR and 56.40-dB SFDR for a 7.7-kHz input. The ADC output rates decrease
to 5 MS/s and 2.5 MS/s for $N_{cT}=2$ and 4. Across all evaluated RF
traces, median NCC values were 0.981 and 0.932, with median NRMSE values
of 0.36 and 0.56, respectively. Wire-phantom localization error remained
below 0.04 mm with no appreciable FWHM degradation. In the speckle-rich
cyst phantom, SSIM remained 0.94 and 0.87, while CNR decreased from 3.534
in the reference to 2.047 and 1.379. These results demonstrate a
hardware-realistic tradeoff in which temporal compression substantially
reduces ADC conversion count and output data rate while preserving
point-target geometry, whereas low-contrast cyst conspicuity is more
compression-sensitive.
\end{abstract}


\section{Introduction}

Medical ultrasound imaging is increasingly moving from conventional
cart-based scanners toward compact, catheter-based, wearable, wireless,
and point-of-care platforms for diagnostic imaging, intravascular
imaging, intracardiac imaging, physiologic monitoring, 3D imaging, and
image-guided intervention \cite{chen2021integrated,giangrossi2022requirements,black1994cmos,gurun2014singlechip,chen2017frontend,rezvanitabar2022integrated,montaldo2009coherent}. These applications require
small probe form factors, low power consumption, reduced cabling, and
high frame-rate operation while preserving sufficient image quality for
tissue visualization and quantitative assessment. This creates a
fundamental receiver-side bottleneck: advanced ultrasound imaging
benefits from many receive elements, high acquisition rates, and access
to pre-beamformed radio-frequency (RF) channel data, whereas
miniaturized biomedical systems are constrained by area, power, memory,
and data-transfer bandwidth.

In a conventional array-based ultrasound system, each active element is
sampled and digitized at a rate sufficient to preserve the RF bandwidth
and support accurate receive delay estimation. The resulting channel
data are then transferred to a backend processor for beamforming,
compounding, Doppler processing, or other reconstruction algorithms.
This architecture preserves maximum imaging flexibility because the raw
pre-beamformed RF data remain available. However, it also imposes a
large hardware and data burden through the analog front-end circuits,
analog-to-digital converters (ADCs), sampling clocks, local storage, and
output links required for each channel \cite{chen2021integrated}, \cite{giangrossi2022requirements}, \cite{rezvanitabar2022integrated}.
Integrated ultrasound systems have therefore used strategies such as
time-division multiplexing, sub-aperture or micro-beamforming, and
sparse-array operation to reduce the number of active receive paths or
transmitted data streams \cite{black1994cmos,gurun2014singlechip,chen2017frontend,rezvanitabar2022integrated}. These approaches have enabled
important progress, but they introduce trade-offs. Time-division
multiplexing can limit effective frame rate, micro-beamforming reduces
access to individual channel data, and sparse apertures can affect
resolution, sidelobes, and robustness unless the array and
reconstruction method are carefully co-designed.

Compressive sensing (CS) offers an alternative acquisition framework in
which a signal that is sparse or compressible in a suitable domain can
be recovered from fewer measurements than required by direct Nyquist
sampling, provided that the measurement process is sufficiently
incoherent with the sparse representation \cite{candes2008introduction,candes2006robust,donoho2006compressed}. CS has been
highly influential in imaging modalities such as MRI and CT, but its
translation to ultrasound is more challenging because ultrasound RF data
contain band-limited pulse echoes, phase-sensitive propagation, coherent
speckle, and strong dependence on aperture geometry and beamforming.
Early CS-related ultrasound studies addressed this challenge by
exploiting structure in pulse-echo signals, including
finite-rate-of-innovation and Xampling-based models \cite{tur2011innovation}, compressed
and Fourier-domain beamforming \cite{wagner2012compressed}, \cite{chernyakova2014fourier}, and pre-beamformed RF
reconstruction using wave atoms, wavelets, or learned dictionaries
\cite{liebgott2013prebeamformed}, \cite{lorintiu2015dictionary}. Other methods targeted the multi-channel nature of
array imaging by compressing or selecting receive channels, thereby
connecting CS to potential reductions in cable count, ADC count, and
receive-chain complexity \cite{besson2017approach,mitrovic2020hardware,anand2021practical,mamistvalov2022convolutional}.

Despite this progress, a gap remains between CS ultrasound theory and
practical integrated receiver hardware. Many prior demonstrations use
simulation, digitally subsampled data acquired by conventional systems,
or compression applied after beamforming. In those cases, the reported
data-reduction factor does not necessarily translate into reduced ADC
activity, memory traffic, or output-link bandwidth at the front-end.
Moreover, if compression is performed only after beamforming or after
irreversible sub-aperture summation, the system may lose access to
pre-beamformed RF data required for dynamic focusing, plane-wave
compounding, Doppler processing, elastography, or future RF-domain
algorithms.

A second important gap is validation on realistic ultrasound scenes.
Wire targets, point reflectors, and sparse absorbers are useful for
verifying timing, localization, and point-spread-function behavior, but
they are favorable cases for CS because the underlying scene is already
sparse or highly compressible. Diagnostic ultrasound, in contrast, often
contains diffuse scattering, coherent speckle, shadowing, extended
structures, and low-contrast lesions. Therefore, a convincing CS
ultrasound receiver should be evaluated not only on sparse wire-like
targets but also on speckle-rich or non-point-target phantoms, such as
cyst phantoms, where lesion contrast, speckle texture, and boundary
preservation can be assessed.

Recent circuit-level work in photoacoustic imaging has shown that CS
operations can be embedded directly into the receiver electronics. Liao
et al. reported a compressive-sensing photoacoustic receiver using a
matrix-vector-multiplication SAR ADC, where programmable ternary weights
were applied to channel signals before summation and digitization
\cite{liao2025mvm}. This work is an important circuit demonstration because it
integrates compression with the ADC rather than treating CS only as an
off-chip algorithm. However, its imaging validation used sparse,
high-contrast photoacoustic targets, and larger apertures were emulated
by mechanical scanning. This leaves open the question of whether
hardware-integrated CS can preserve clinically relevant information in
diffuse, speckle-rich, or non-sparse ultrasound imaging scenarios.

This paper presents a compressed-sensing medical ultrasound receiver
based on a pipelined SAR ADC architecture that performs temporal compression and
sub-Nyquist sampling during RF acquisition. The proposed approach embeds
pseudo-random polarity modulation and multi-sampling into the pipelined SAR ADC
sampling network, so that multiple consecutive RF samples are
accumulated in the charge domain and digitized using a single
conversion, enabling temporal compression ratios of $1\times$, $2\times$, and $4\times$.
Unlike post-acquisition digital compression, this strategy reduces SAR
conversion activity and ADC output samples at the point of measurement.
The compressed measurements are then used to recover pre-beamformed RF
channel data, preserving compatibility with conventional ultrasound
image-formation pipelines. The work is validated using measured ASIC
outputs, RF-domain recovery, and both wire and cyst phantom imaging,
with the cyst phantom providing the key evidence that temporal CS
acquisition can preserve meaningful image structure beyond sparse
point-target scenes. Together, these results connect an in-ADC CS
circuit mechanism to RF recovery and B-mode image formation, helping
bridge the gap between CS theory and practical ultrasound receiver
design. To the best of our knowledge, this is the first on-chip temporal
compressive acquisition approach for medical ultrasound RF imaging that
demonstrates sub-Nyquist RF recovery from measured ASIC outputs and
validates B-mode reconstruction on a speckle-rich cyst phantom, rather
than only sparse point-like targets.

The remainder of this manuscript is organized as follows. Section II
describes the temporal CS acquisition model and RF-domain recovery
formulation. Section III presents the CS-pipelined SAR ADC architecture and
circuit implementation. Section IV describes the RF recovery and
beamforming pipeline. Section V reports measured ADC performance,
waveform and RF-line recovery, and wire and cyst phantom imaging
results. Section VI discusses limitations, benchmarking, and future
extensions toward space-time compression for integrated ultrasound
systems.

\section{Temporal CS Acquisition Approach}

The objective of the proposed acquisition approach is to reduce the
number of digitized samples and the output data rate of pre-beamformed
ultrasound RF data without discarding the receive-aperture information
that is required for conventional beamforming. In standard ultrasound
receive chains, each active element is amplified, filtered, sampled, and
digitized at a rate selected to preserve the RF bandwidth and to support
accurate receive delays. For an RF line, a conventional ADC produces
\(N_{L}\) digital values per channel and per firing. In
high-channel-count, multiple plane-wave angles, or high-frame-rate
acquisitions, this sampling burden directly translates into ADC
conversion activity, memory traffic, digital I/O bandwidth, and backend
processing load.

In the proposed temporal CS-SAR acquisition, compression is performed
during sampling rather than after full-rate digitization. Multiple
consecutive RF samples are multiplied by known pseudo-random polarity
coefficients, accumulated in the pipelined SAR ADC sampling network, and digitized
using one conversion. Therefore, the ADC output is a coded measurement
of a short RF segment, not a decimated subset of the original RF
waveform. The present prototype demonstrates temporal compression only;
all receive channels are preserved and reconstructed before beamforming.

\begin{figure*}[!t]
\centering
\includegraphics[width=0.96\textwidth]{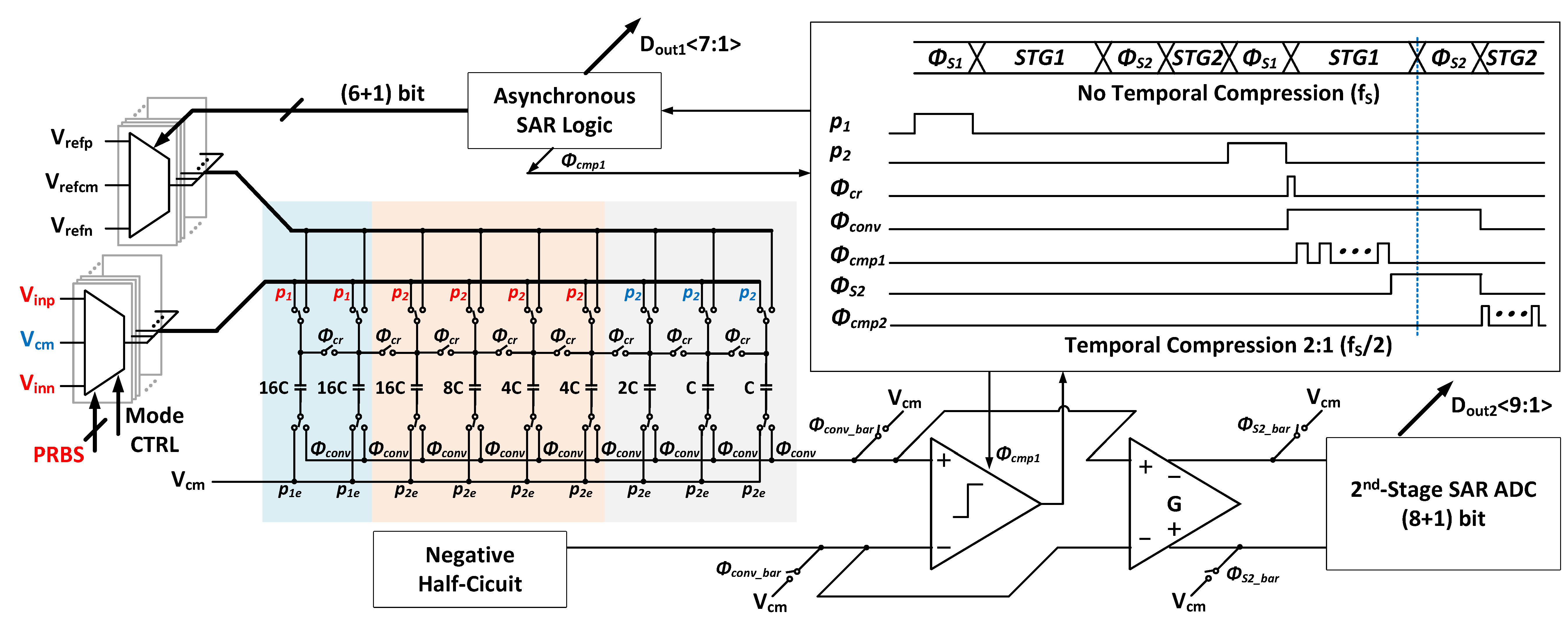}
\caption{Architecture of the proposed CS-pipelined SAR ADC. The first-stage CDAC performs PRBS-controlled polarity selection, capacitor-subset sampling, and charge-domain accumulation before SAR quantization. The residue is amplified by an interstage gain of 16 and digitized by the second-stage SAR ADC.}
\label{fig:architecture}
\end{figure*}

\subsection{Temporal Measurement Model}

For one receive element and one transmit event, let \(x_{i}\) denote the
uncompressed pre-beamformed RF vector from element i,

\begin{equation}
\mathbf{x}_{i}=\big[x_i[0],x_i[1],\ldots,x_i[N_L-1]\big]^{T}\in\mathbb{R}^{N_L}.
\label{eq:xi}
\end{equation}

Here, \(N_{L}\) is the number of RF samples that would be acquired in
the corresponding non-compressed mode. The non-compressed mode
corresponds to $N_{cT}=1$, where the ADC output represents the
full RF vector $\mathbf{x}_i$, up to quantization and analog front-end
errors. In temporal CS mode, the ADC produces a shorter measurement
vector $\mathbf{y}_{i} \in \mathbb{R}^{M}$, with $M$ samples, where

\begin{equation}
M=\frac{N_L}{N_{cT}},\qquad N_{cT}\in\{1,2,4\}.
\label{eq:M}
\end{equation}

Here, $N_{cT}$ is the temporal compression ratio. For
$N_{cT}=2$, two nominal RF samples are encoded into one compressed
measurement before one SAR conversion. For $N_{cT}=4$, four
nominal RF samples are encoded into one compressed measurement. If
required, $N_L$ is padded or truncated so that it is divisible by
$N_{cT}$.

The temporal compressed measurement is modeled as

\begin{equation}
\mathbf{y}_i=\mathbf{C}_T\mathbf{x}_i,\qquad \mathbf{C}_T\in\mathbb{R}^{M\times N_L}.
\label{eq:temporal_measurement}
\end{equation}

where $\mathbf{C}_T$ is the temporal sensing matrix implemented by
PRBS-controlled polarity selection and charge-domain accumulation in the
CS-SAR ADC. Using zero-based indexing, the $m$-th row of
$\mathbf{C}_T$ contains $N_{cT}$ nonzero entries corresponding to the
samples in the $m$-th compression window:

\begin{equation}
[\mathbf{C}_T]_{m,n}=\begin{cases}
\beta p_{m,r}, & n=mN_{cT}+r,\\
0, & \text{otherwise}.
\end{cases}
\label{eq:CT_entries}
\end{equation}

where $r=0,1,\ldots,N_{cT}-1$. Here, (\(p_{\{ m,r\}} \in \{ - 1, + 1\}\)) is the PRBS polarity applied
to the $r$-th sample in the $m$-th temporal compression
window, and (\(\beta\)) is an effective scaling factor that accounts for
normalization and charge-domain weighting in the ADC. For an ideal
equal-weight model, (\(\beta\)) can be absorbed into the digital scaling
of $\mathbf{y}_i$; in the measured circuit, it represents the effective
gain of the implemented sampling and charge-redistribution network.

For $N_{cT}=1$, the temporal sensing matrix reduces to the
identity matrix,

\begin{equation}
\mathbf{C}_T=\mathbf{I},\qquad \mathbf{y}_i=\mathbf{x}_i.
\label{eq:nct1}
\end{equation}

For $N_{cT}=2$, each ADC conversion produces a signed combination
of two consecutive RF samples,

\begin{equation}
y_i[m]=\beta\!\left(p_{m,0}x_i[2m]+p_{m,1}x_i[2m+1]\right).
\label{eq:nct2}
\end{equation}

For $N_{cT}=4$, each ADC conversion combines four consecutive RF
samples,

\begin{equation}
\begin{split}
y_i[m]=\beta\big(&p_{m,0}x_i[4m]+p_{m,1}x_i[4m+1]\\
&+p_{m,2}x_i[4m+2]+p_{m,3}x_i[4m+3]\big).
\end{split}
\label{eq:nct4}
\end{equation}

Equations~\eqref{eq:nct2} and~\eqref{eq:nct4} describe the core temporal CS operation. The ADC
does not simply discard samples; instead, it embeds multiple RF samples
into a lower-dimensional set of coded measurements before quantization.

\subsection{Extension to the Receive Aperture}

For an array with $N_e$ receive elements, the uncompressed RF data
for one transmit event can be written as

\begin{equation}
\mathbf{x}=\big[\mathbf{x}_1^T,\mathbf{x}_2^T,\ldots,\mathbf{x}_{N_e}^T\big]^T
\in\mathbb{R}^{N_eN_L}.
\label{eq:array_x}
\end{equation}

Because the present implementation performs temporal compression only,
the measurement operator does not mix different receive channels. The
array-level measurement model is therefore block diagonal:

\begin{equation}
\begin{split}
\mathbf{y}&=\mathbf{C}_{\mathrm{blk}}\mathbf{x},\\
\mathbf{C}_{\mathrm{blk}}&=\operatorname{blockdiag}\!\left(\mathbf{C}_{T,1},\ldots,\mathbf{C}_{T,N_e}\right).
\end{split}
\label{eq:blockdiag}
\end{equation}

The same temporal sensing pattern may be used for all receive channels,
or channel-dependent PRBS patterns may be assigned so that
$\mathbf{C}_{T,i}$ differs between elements. In either case, the receive
aperture is preserved after RF recovery. Therefore, the demonstrated
hardware reduction in this prototype is a reduction in SAR conversion
events and ADC output samples by the factor $N_{cT}$, not a
reduction in receive-channel count.

\begin{figure*}[!ht]
\centering
\subfloat[]{\includegraphics[width=0.45\textwidth]{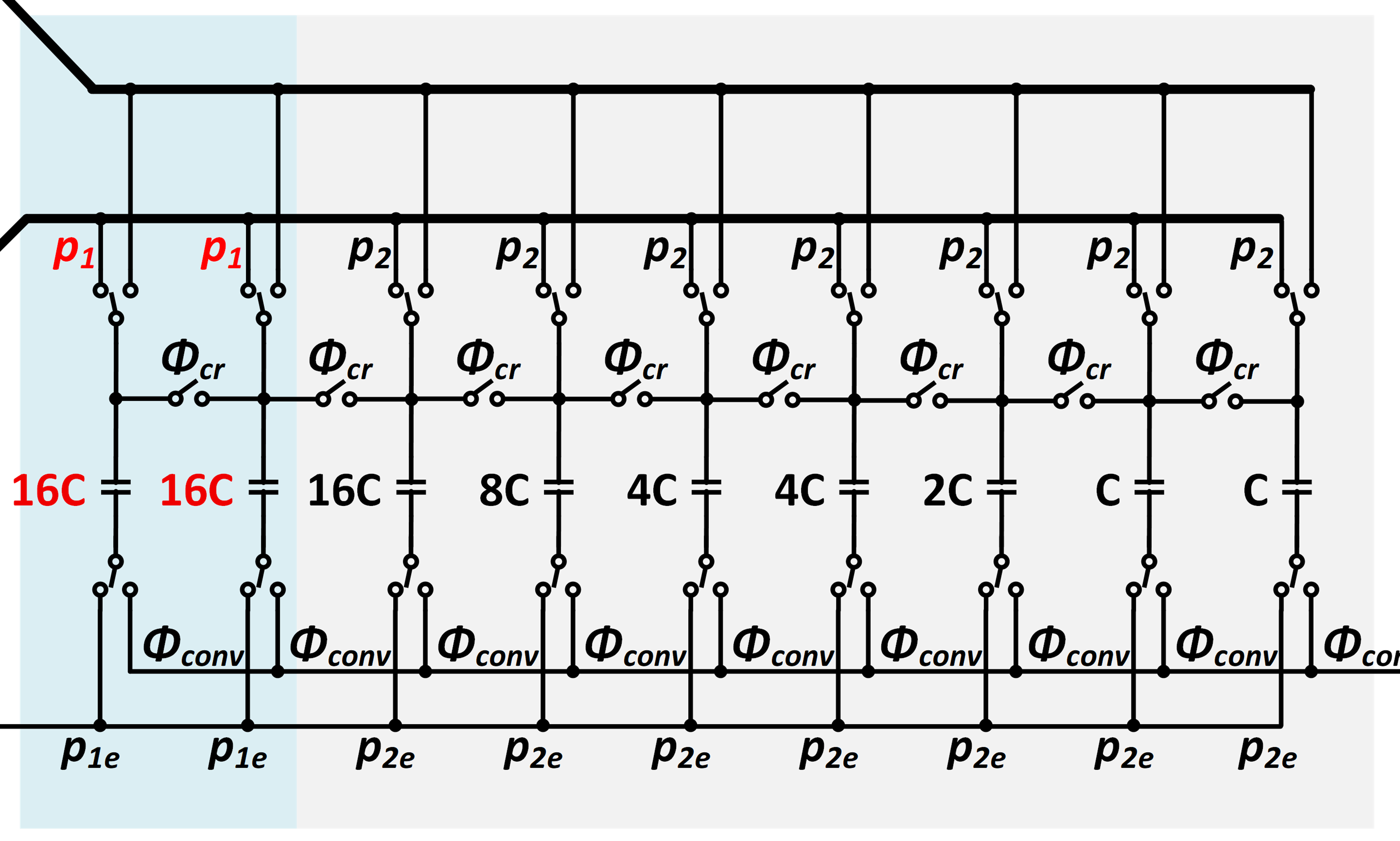}}\hfill
\subfloat[]{\includegraphics[width=0.45\textwidth]{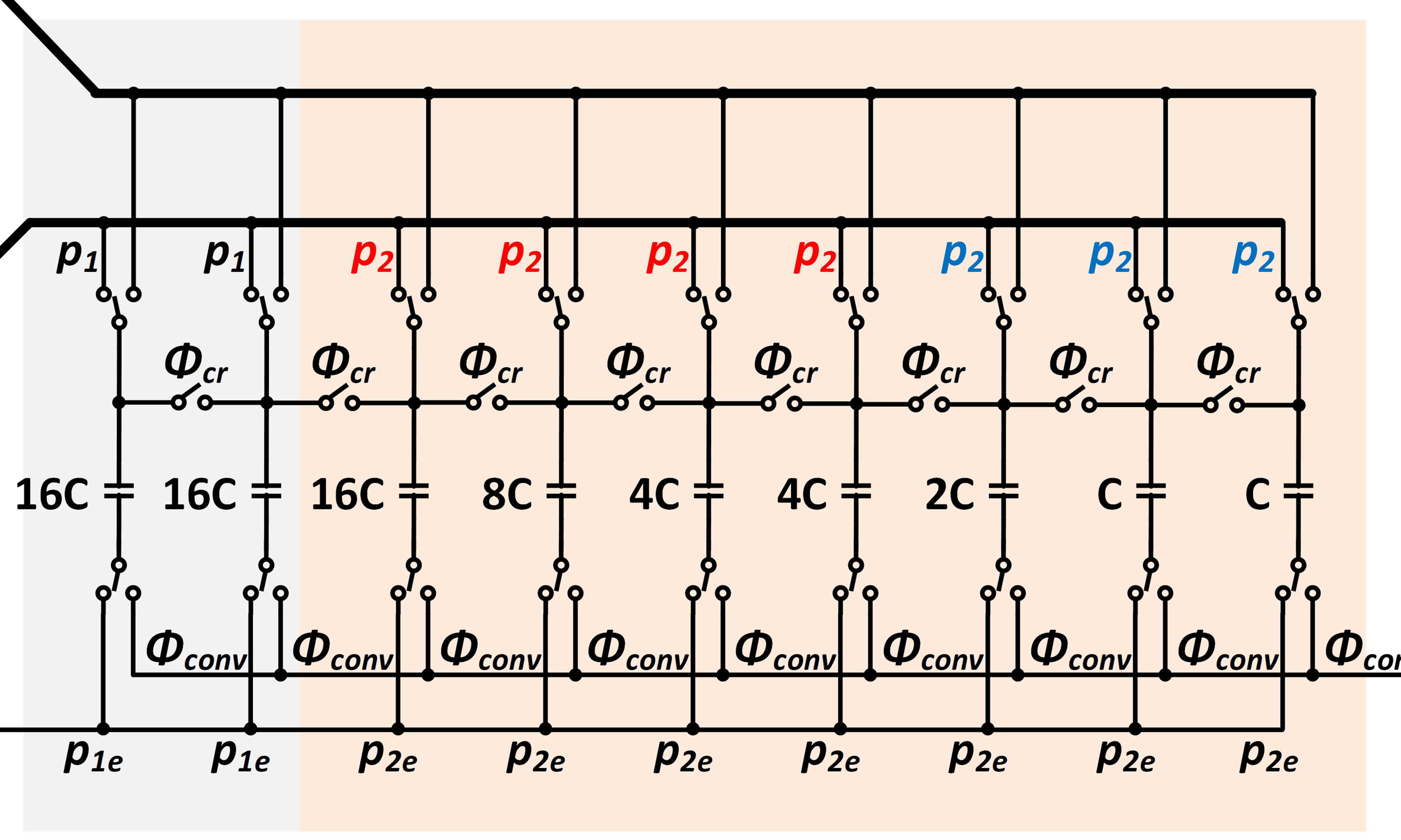}}\hfill
\subfloat[]{\includegraphics[width=0.45\textwidth]{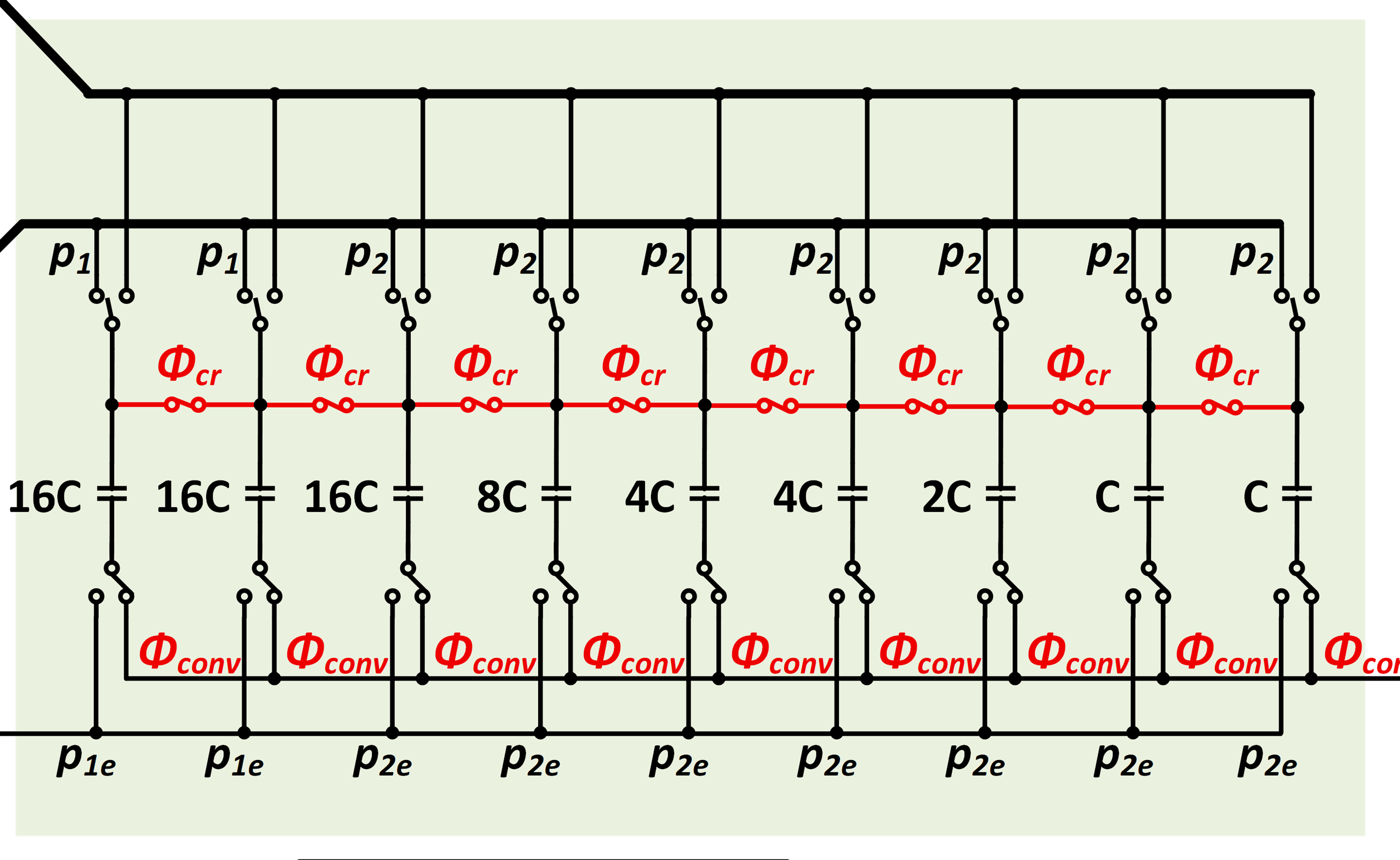}}\hfill
\subfloat[]{\includegraphics[width=0.45\textwidth]{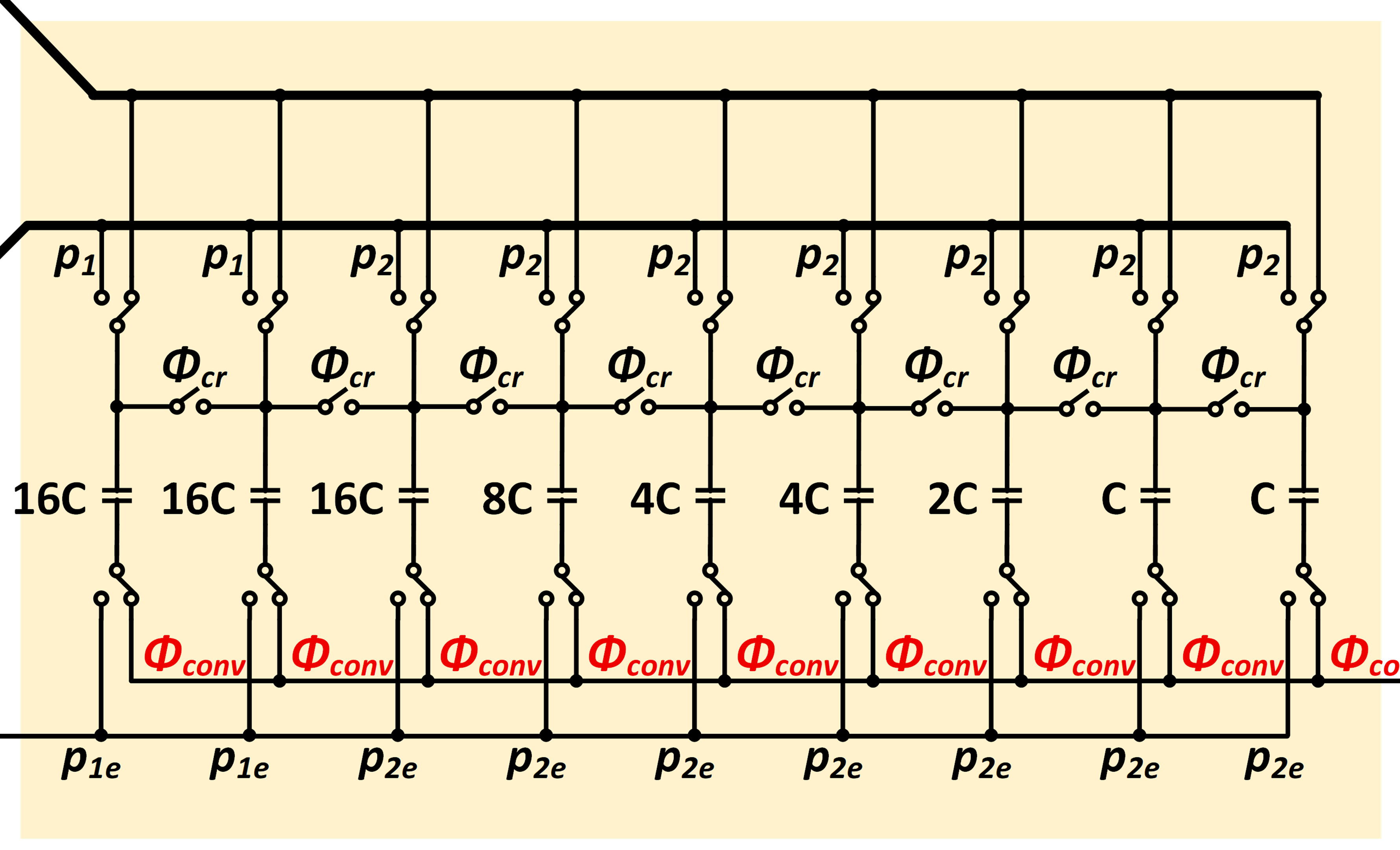}}\hfill
\caption{Step-by-step operation of charge-domain temporal compression in the first-stage CDAC, illustrated for $N_{cT}=2$. (a) During the first sampling subphase, a selected capacitor subset samples the first RF value with PRBS-controlled differential polarity $p_1$. (b) During the second sampling subphase, the capacitor subset assigned to the next sample acquires the second RF value with polarity $p_2$, while the charge stored from the first subphase is retained. (c) After both samples have been acquired, $\phi_{\mathrm{cr}}$ is asserted and charge redistribution combines the stored capacitor charges into a single PRBS-weighted analog measurement. (d) The redistribution phase is released and $\phi_{\mathrm{conv}}$ initiates SAR quantization of the compressed voltage; the resulting residue is subsequently amplified and digitized by the second stage as shown in Fig.~\ref{fig:architecture}. Thus, two sampling instants on the nominal RF sampling grid produce one ADC output conversion in the $N_{cT}=2$ mode. For $N_{cT}=4$, the same operation is extended to four PRBS-weighted sampling subphases before charge redistribution and conversion.}
\label{fig:charge_compression}
\end{figure*}

\subsection{RF-Domain Sparse Representation}

The RF recovery model uses a pulse-dictionary representation rather than
assuming sparsity of the final B-mode image. This distinction is
important because medical ultrasound images often contain speckle,
tissue texture, low-contrast boundaries, and anechoic or hypoechoic
inclusions. These features are not generally sparse in the displayed
image domain. Instead, each pre-beamformed RF trace is represented as a
superposition of delayed copies of the system pulse response, following
a physics-matched RF-domain model.

For one receive channel,

\begin{equation}
\mathbf{x}_i=\mathbf{H}_i\mathbf{c}_i.
\label{eq:pulse_model}
\end{equation}

where $\mathbf{H}_i$ is a pulse-dictionary or RF forward matrix and
$\mathbf{c}_i$ is the coefficient vector to be estimated. Each column of
$\mathbf{H}_i$ corresponds to a delayed copy of an ultrasound pulse
$h$. The pulse $h$ can be obtained from a measured wire or
point-reflector response using the same probe and acquisition chain, or
from the Field II impulse response used in the controlled simulation
experiments.

A one-channel pulse dictionary can be written as a lower-triangular
Toeplitz-like convolutional matrix,

\begin{equation}
\mathbf{H}_i=\begin{bmatrix}
h_0 & 0 & 0 & \cdots & 0\\
h_1 & h_0 & 0 & \cdots & 0\\
h_2 & h_1 & h_0 & \cdots & 0\\
\vdots & \vdots & \vdots & \ddots & \vdots\\
0 & \cdots & h_2 & h_1 & h_0
\end{bmatrix}.
\label{eq:Hi}
\end{equation}

This representation imposes sparsity or compressibility on the RF
echo-generation process, not directly on the displayed B-mode image. It
is therefore more suitable for speckle-rich or cyst-phantom imaging than
a model that relies only on point-like image sparsity.

For the full receive aperture, the RF-domain model can be written as

\begin{equation}
\mathbf{x}=\mathbf{H}\mathbf{c},\qquad
\mathbf{H}=\begin{bmatrix}
\mathbf{H}_1 & 0 & \cdots & 0\\
0 & \mathbf{H}_2 & \cdots & 0\\
\vdots & \vdots & \ddots & \vdots\\
0 & 0 & \cdots & \mathbf{H}_{N_e}
\end{bmatrix}.
\label{eq:array_H}
\end{equation}

In the present implementation, a shared or calibrated average pulse
dictionary is used as the baseline model. Channel-specific dictionaries
can be incorporated if element responses or analog front-end
characteristics vary significantly across the array.

\subsection{RF Recovery From Compressed Measurements}

Substituting the RF-domain model in~\eqref{eq:pulse_model} into the temporal measurement
model in~\eqref{eq:temporal_measurement} gives

\begin{equation}
\mathbf{y}_i=\mathbf{C}_T\mathbf{H}_i\mathbf{c}_i
=\mathbf{A}_i\mathbf{c}_i,\qquad \mathbf{A}_i=\mathbf{C}_T\mathbf{H}_i.
\label{eq:compressed_dictionary}
\end{equation}

The coefficient vector $\mathbf{c}_i$ is estimated from the compressed ADC
measurements $\mathbf{y}_i$. Because $M<N_L$ for $N_{cT}>1$,
the inverse problem is underdetermined and requires regularization. In
this work, the recovered coefficient vector is obtained using an
elastic-net formulation,

\begin{equation}
\widehat{\mathbf{c}}_i=\arg\min_{\mathbf{c}}
\left\{\left\|\mathbf{y}_i-\mathbf{A}_i\mathbf{c}\right\|_2^2
+\lambda P_{\alpha}(\mathbf{c})\right\}.
\label{eq:elastic_net}
\end{equation}

\begin{equation}
P_{\alpha}(\mathbf{c})=\frac{1}{2}(1-\alpha)\|\mathbf{c}\|_2^2
+\alpha\|\mathbf{c}\|_1,\qquad 0\leq\alpha\leq1.
\label{eq:elastic_penalty}
\end{equation}

The ($\ell_1$) term promotes sparsity in the delayed-pulse
coefficient domain, while the ($\ell_2$) term stabilizes the
solution and discourages excessive suppression of distributed
low-amplitude echoes. This is useful for cyst and speckle-rich phantoms,
where a purely sparse solution may preserve strong reflectors but
distort weak scatterers or speckle texture.

After estimating (\({\widehat{c}}_{i}\)), the recovered RF vector is
reconstructed as

\begin{equation}
\widehat{\mathbf{x}}_i=\mathbf{H}_i\widehat{\mathbf{c}}_i.
\label{eq:rf_reconstruction}
\end{equation}

The recovered pre-beamformed RF data from all receive elements are then
passed to the same downstream imaging chain used for the non-compressed
reference, including receive-delay calculation, dynamic focusing or
plane-wave beamforming, compounding when multiple transmit angles are
used, envelope detection, log compression, and B-mode display.

\begin{figure}[!t]
\centering
\includegraphics[width=0.88\columnwidth]{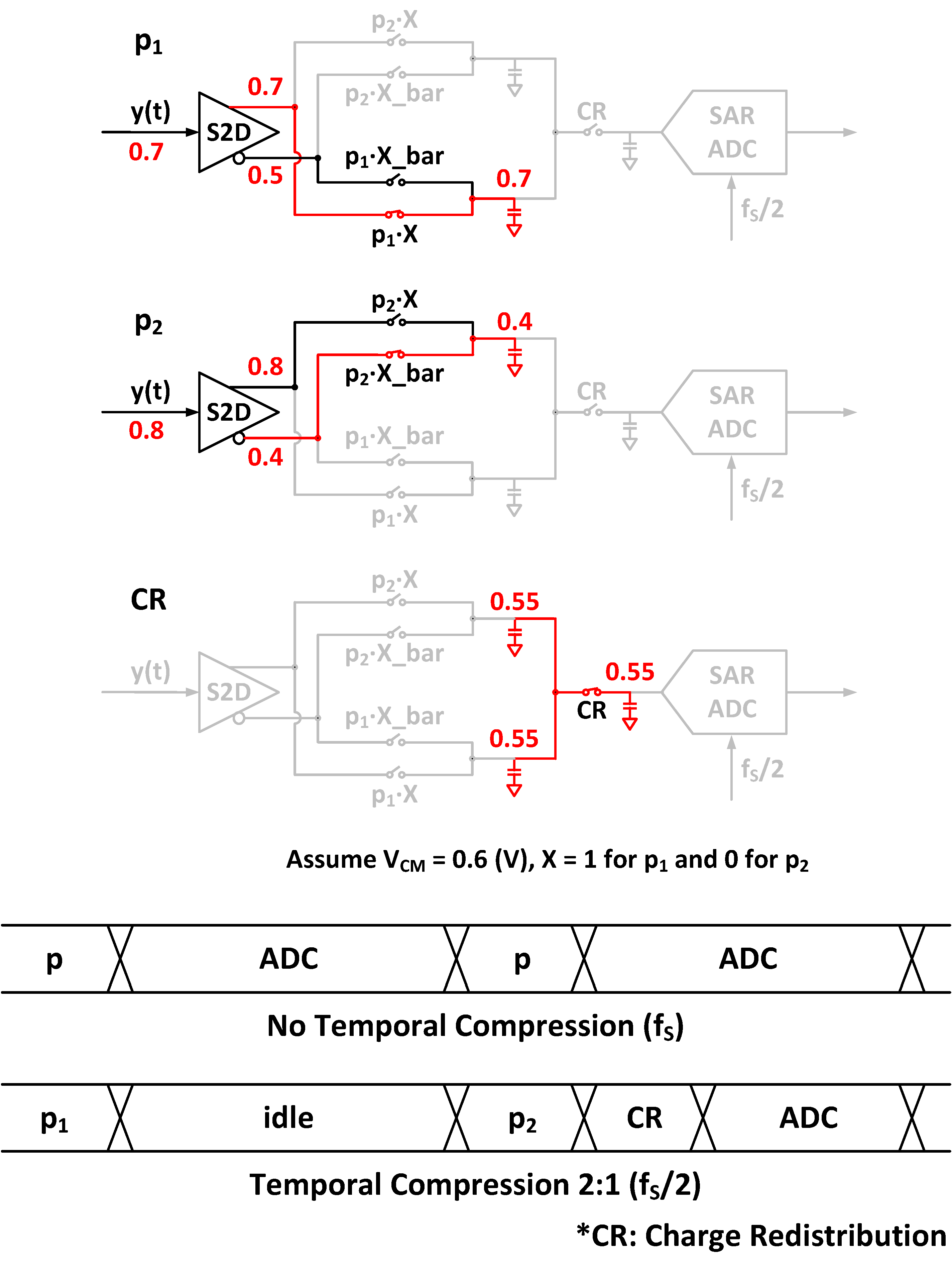}
\caption{Timing illustration of the CS-SAR ADC for the non-compressed
($N_{cT}=1$) and 2$\times$ temporal-compression ($N_{cT}=2$) modes.
In the non-compressed mode, each nominal RF sampling event is followed
by one SAR conversion. In the $N_{cT}=2$ mode, two PRBS-weighted
sampling subphases are followed by charge redistribution and a single
SAR conversion, reducing the ADC output rate from $f_s$ to $f_s/2$.
The $N_{cT}=4$ mode extends the same operation to four PRBS-weighted
sampling subphases before charge redistribution and conversion,
resulting in an output rate of $f_s/4$.}
\label{fig:timing}
\end{figure}

\subsection{Circuit-Level Interpretation}

The temporal sensing matrix $\mathbf{C}_T$ is implemented in the proposed
ADC through the SAR sampling operation. In a conventional SAR ADC, one
RF sample is stored on the sampling capacitor array and then quantized.
In the proposed CS-SAR ADC, multiple consecutive RF samples are first
sign-modulated by PRBS-controlled differential polarity selection and
stored on selected capacitor subsets. After the $N_{cT}$ sampling
subphases, charge redistribution forms a weighted analog sum, and a
single SAR conversion digitizes the compressed measurement.

For a differential RF input, multiplication by (+1) or (-1) is
implemented by selecting the normal or inverted input polarity during
the corresponding sampling subphase. Thus, the multiply-and-accumulate
operation in~\eqref{eq:nct2} and~\eqref{eq:nct4} is performed before quantization without
requiring an explicit analog multiplier. The ADC output rate is reduced
from $f_s$ to (\(f_{s}/N_{cT}\)), where $f_s$ is the nominal
RF sampling rate of the non-compressed mode.

This hardware interpretation defines the scope of the present prototype.
The demonstrated CS-SAR ADC reduces temporal conversion count and output
data volume while preserving all receive channels for reconstruction and
beamforming. Spatial channel compression, which would reduce the number
of receive paths or ADC instances, is a future extension rather than a
claim of the present measured implementation.

\section{CS-SAR ADC and Circuit Implementation}

The temporal CS model described in Section II requires a receiver
circuit that can form known signed combinations of consecutive RF
samples before digitization. In the proposed implementation, this
operation is embedded directly into the sampling network of a SAR ADC.
During each compression window, multiple RF samples are selected with
pseudo-random differential polarity, stored on capacitor subsets,
combined through charge redistribution, and digitized using one SAR
conversion. The ADC therefore acts as an analog-to-information
converter, producing coded temporal measurements rather than uniformly
sampled RF values.

The prototype supports three operating modes: $N_{cT}=1$,
corresponding to conventional non-compressed acquisition;
$N_{cT}=2$, in which two PRBS-weighted RF samples are accumulated
before one conversion; and $N_{cT}=4$, in which four PRBS-weighted
RF samples are accumulated before one conversion. The receive aperture
is unchanged in all modes. Thus, the demonstrated hardware reduction is
in temporal conversion count and ADC output data rate, while RF recovery
and beamforming are performed off chip.

\subsection{Architecture Overview}

Fig.~\ref{fig:architecture} shows the architecture of the proposed CS-SAR ADC. The
converter uses a two-stage pipelined SAR structure. The first stage
performs both coarse quantization and temporal CS sampling using a
reconfigurable capacitive DAC (CDAC), while the second stage digitizes
the amplified residue using a conventional SAR conversion. The main
signal path consists of PRBS-controlled differential input-polarity
selection, first-stage CDAC subset sampling, charge redistribution,
first-stage SAR decision, residue amplification, second-stage SAR
conversion, and digital output generation.

The first-stage sampling network implements the temporal sensing matrix
$\mathbf{C}_T$. Before each sampling subphase, the PRBS bit selects either
the normal or inverted differential input polarity. This realizes
multiplication by (+1) or (-1) without an explicit analog multiplier.
After $N_{cT}$ sampling subphases, the stored charges are
redistributed to form one compressed analog value. This value is then
quantized through the two-stage SAR conversion sequence.

A two-stage SAR topology is used to reduce the first-stage CDAC size
while maintaining sufficient resolution for ultrasound RF acquisition.
The first stage produces a residue voltage after the coarse SAR
decision, and the residue is amplified by an inter-stage gain of 16
before second-stage quantization. Redundancy is included between the two
stages to improve tolerance to comparator error, residue-amplifier gain
error, and incomplete settling.

\begin{figure*}[!t]
\centering
\subfloat[Nonoverlapping clock generator.]{\includegraphics[width=0.30\textwidth]{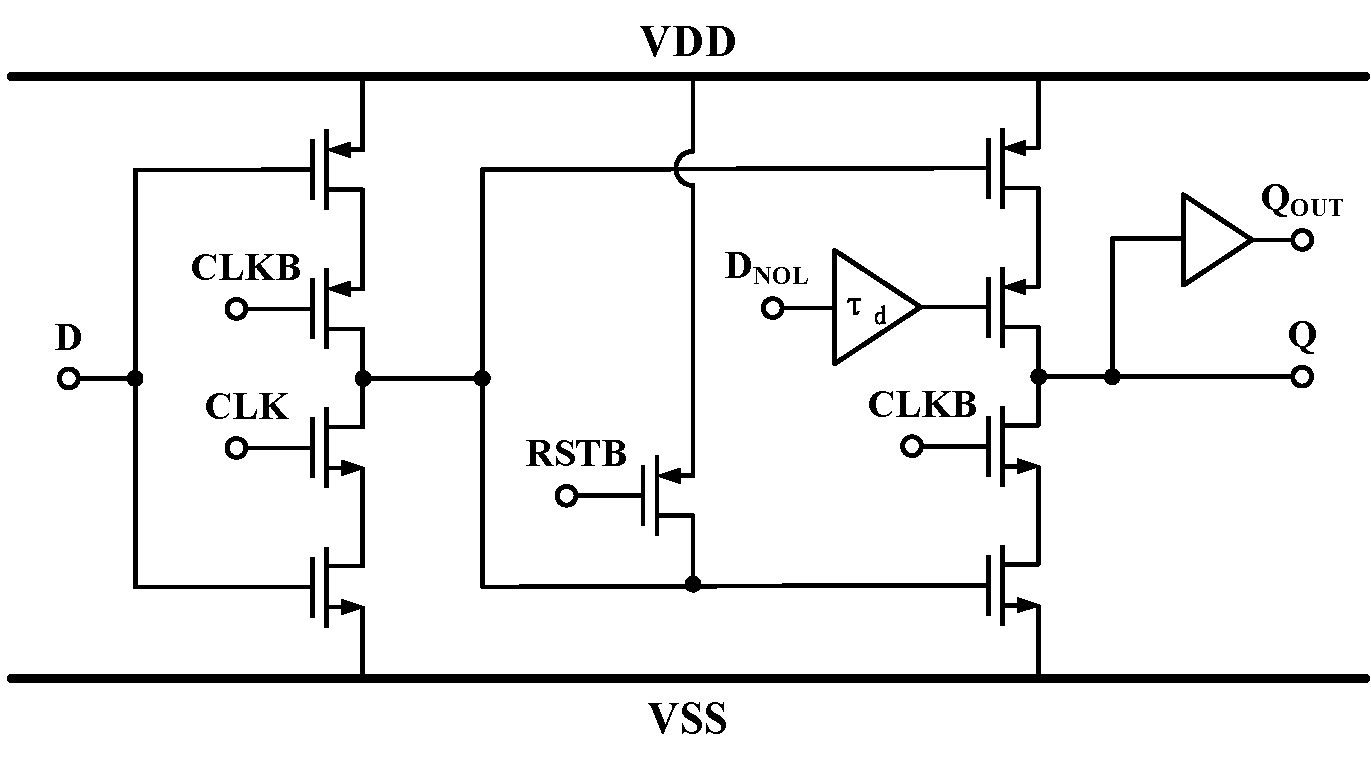}\label{fig:clockgen}}
\hfil
\subfloat[StrongARM latch comparator.]{\includegraphics[width=0.27\textwidth]{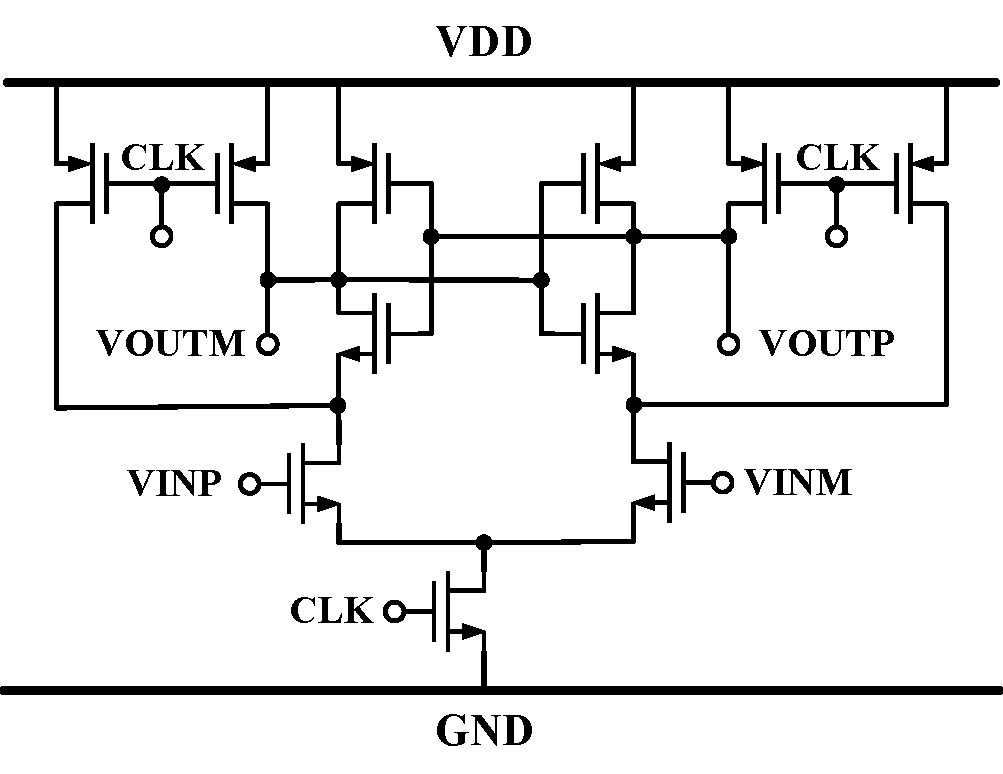}\label{fig:strongarm}}
\hfil
\subfloat[Inverter-based interstage residue amplifier.]{\includegraphics[width=0.36\textwidth]{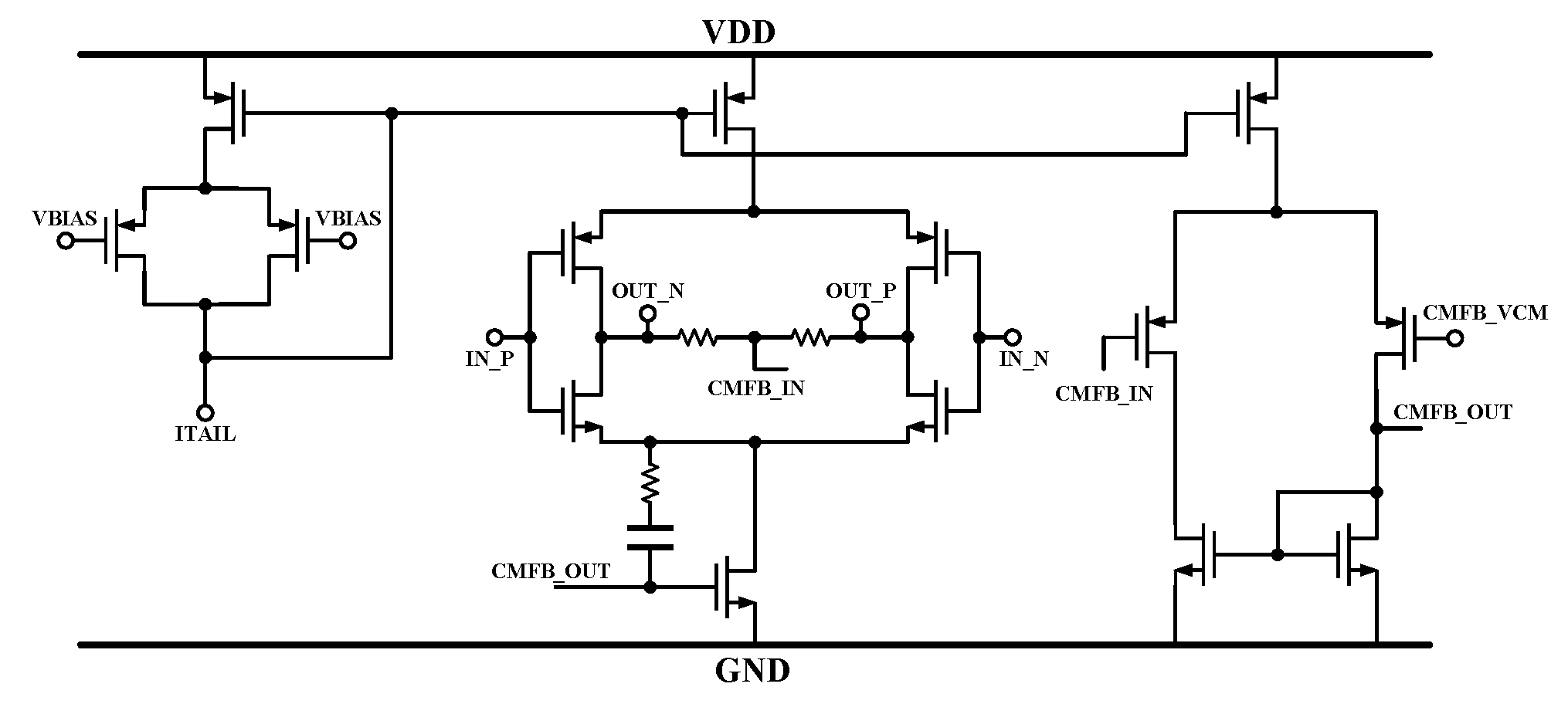}\label{fig:residue_amp}}
\caption{Key circuit building blocks of the proposed CS-SAR ADC.}
\label{fig:building_blocks}
\end{figure*}

\subsection{Charge-Domain Temporal Compression}

The circuit operation can be related directly to the temporal
measurement model in Section II. Let (\(v_{in}\lbrack k\rbrack\)) denote
the differential RF input sample at the $k$-th nominal sampling
instant. For the $m$-th compression window, the ideal compressed
voltage generated by charge redistribution can be written as

\begin{equation}
v_{\mathrm{CS}}[m]=\gamma\sum_{r=0}^{N_{cT}-1}p_{m,r}
 v_{\mathrm{in}}[mN_{cT}+r]+v_{\mathrm{err}}[m].
\label{eq:charge_domain}
\end{equation}

where (\(p_{\left\{ m,r \right\}} \in \left\{ - 1, + 1 \right\}\)) is
the PRBS polarity applied during the $r$-th sampling subphase,
(\(\gamma\)) is the effective charge-domain gain, and
(\(v_{err}\lbrack m\rbrack\)) represents circuit nonidealities such as
capacitor mismatch, finite settling, charge injection, clock
feedthrough, leakage, comparator kickback, and residue-amplifier error.
The factor (\(\gamma\)) corresponds to the circuit-level scaling
represented by (\(\beta\)) in Section II and can be included in the
effective sensing matrix used for reconstruction.

In the $N_{cT}=2$ or $N_{cT}=4$ modes, two or four
consecutive RF samples are sampled onto selected first-stage CDAC
subsets with independently assigned PRBS polarities and are then
combined before charge redistribution and SAR conversion. The
implemented CDAC partitioning assigns approximately half of the
available sampling capacitance to each sample in the $2\times$ mode and
approximately one quarter to each sample in the $4\times$ mode, with redundancy
and common-mode-connected capacitance included in the effective gain
calibration.

This operation differs from ordinary downsampling. A downsampled ADC
output would retain only one sample from each time window and discard
the others. In contrast, the proposed CS-SAR ADC stores a known signed
combination of all samples in the window before quantization. The
missing RF samples are therefore not simply omitted; they are encoded
into coded measurements and recovered digitally using the known PRBS
sequence and RF pulse-dictionary model.

\subsection{Compression Modes and Timing}

The nominal non-compressed mode operates at an RF sampling rate of 10
MS/s and corresponds to $N_{cT}=1$. In this mode, each RF sample
produces one ADC output code. In the 2x temporal compression mode, two
PRBS-weighted sampling subphases are performed before one conversion,
reducing the effective ADC output rate to 5 MS/s. In the 4x temporal
compression mode, four PRBS-weighted sampling subphases are performed
before one conversion, reducing the effective ADC output rate to 2.5
MS/s.

The timing is generated from a 50-MHz master clock, corresponding to a
20-ns timing resolution. In the non-compressed mode, the ADC operates
with a 100-ns nominal conversion period. The first-stage sampling,
charge redistribution, first-stage SAR decision, residue amplification,
second-stage sampling, and second-stage conversion phases are scheduled
within this clocking framework. In the compression modes, the
first-stage sampling opportunities remain aligned to the nominal RF
sampling grid, while the second-stage conversion is activated only after
the required number of PRBS-weighted sampling subphases has been
accumulated.

Maintaining the same nominal RF sampling grid simplifies comparison
across compression modes. The $N_{cT}=1$ mode provides the
measured ASIC reference, while the $N_{cT}=2$ and $N_{cT}=4$
modes reduce the number of conversion events and output codes for the
same RF acquisition window.

\begin{figure}[!t]
\centering
\includegraphics[width=0.86\columnwidth]{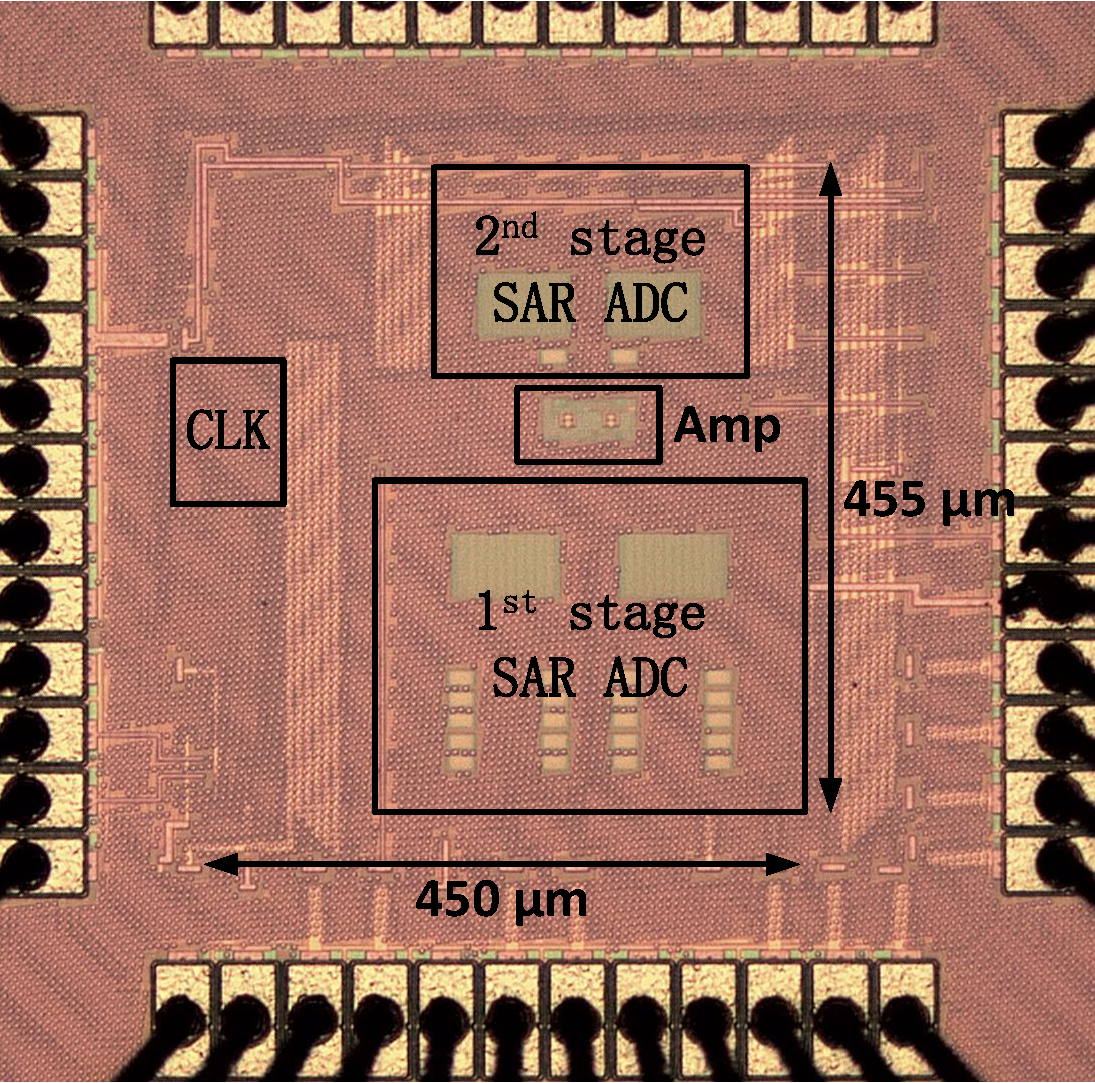}
\caption{Die micrograph of the fabricated CS-SAR ADC. The chip was implemented in 65-nm CMOS and measured using a 1.2-V supply, 1.2-V reference, and 50-MHz master clock.}
\label{fig:die}
\end{figure}

\subsection{Circuit Building Blocks}

\subsubsection{PRBS Polarity Selector and CDAC Sampling Network}

The PRBS polarity selector is implemented using switches that connect
the selected sampling capacitor subset to either the normal or inverted
differential input. This realizes the signed coefficients required by
the temporal CS measurement matrix with small digital and switching
overhead. Because the polarity operation is performed through
differential input selection, no active analog multiplier is required.

The first-stage CDAC is partitioned into capacitor subsets assigned to
different sampling subphases. Each subset stores one PRBS-weighted RF
sample during the compression window. After the final subphase, the
charge-redistribution phase combines the stored charges and presents the
resulting compressed voltage to the first-stage SAR comparator.
Bottom-plate sampling is used to reduce signal-dependent charge
injection and improve sampling linearity.

\subsubsection{Nonoverlapping Clock Generator}

The sampling, charge-redistribution, SAR decision,
residue-amplification, and second-stage conversion phases are generated
by a non-overlapping clock generator driven by the 50-MHz master clock.
The clocking network provides the control phases required for the three
operating modes and prevents overlap between sampling and conversion
operations. In compression modes, the timing generator preserves the
first-stage sampling instants and suppresses unnecessary second-stage
conversion events until the selected compression window has been
accumulated.

This timing structure is important because the circuit must satisfy two
requirements simultaneously: the input samples must be acquired on the
nominal RF sampling grid, and the compressed value must settle
sufficiently before the SAR decision sequence. The charge-redistribution
phase therefore defines the interface between the analog CS accumulation
operation and the quantization process.

\subsubsection{StrongARM Comparator}

Both SAR stages use StrongARM latch comparators. During reset, the
internal and output nodes are precharged to remove memory from the
previous comparison. During regeneration, the differential input
discharges the internal nodes asymmetrically, and the cross-coupled
latch produces a rail-to-rail decision. This comparator topology is
suitable for the prototype because it provides high-speed operation,
compact area, and zero static latch current.

\subsubsection{Interstage Residue Amplifier}

The residue amplifier provides the gain between the first- and
second-stage SAR ADCs. An inverter-based differential amplifier is used
with a nominal inter-stage gain of 16. Complementary NMOS and PMOS input
devices improve transconductance efficiency, while common-mode feedback
stabilizes the output common-mode voltage. The residue amplifier allows
the second stage to resolve the remaining quantization residue and is
therefore a critical block for the linearity and noise performance of
the two-stage converter.

\begin{table}[!t]
\caption{Prototype Implementation Summary}
\label{tab:prototype}
\centering
\footnotesize
\begin{tabular}{@{}ll@{}}
\toprule
Parameter & Value\\
\midrule
CMOS process & 65 nm\\
ADC architecture & Two-stage CS-SAR / pipelined SAR\\
Supported compression modes & $N_{cT}=1,2,4$\\
Nominal RF sampling rate & 10 MS/s\\
Compressed output rates & 5 MS/s ($N_{cT}=2$); 2.5 MS/s ($N_{cT}=4$)\\
Master clock & 50 MHz\\
Supply / reference & 1.2 V / 1.2 V\\
Interstage gain & 16\\
Core area & $450~\mu\mathrm{m}\times455~\mu\mathrm{m}$\\
\bottomrule
\end{tabular}
\end{table}

\subsection{Prototype Implementation Parameters}

The proof-of-concept ADC was fabricated in 65-nm CMOS. The ADC core
occupies approximately (\(450\ \mu\text{m} \times 455\ \mu\text{m}\))
and includes the clock-generation circuitry, first-stage SAR ADC,
inter-stage residue amplifier, and second-stage SAR ADC. The prototype
was measured with a 1.2-V supply, a 1.2-V reference voltage, and a
50-MHz master clock.

In nominal operation, the ADC provides a 10-MS/s output rate. The 2x and
4x temporal compression modes reduce the output rates to 5 MS/s and 2.5
MS/s, respectively. These rates correspond to reduced SAR conversion
activity and reduced output data volume for the same RF acquisition
window. The measured converter performance, including power, SNDR, SFDR,
and imaging validation results, is reported in Section V.

\begin{figure}[!t]
\centering
\includegraphics[width=0.98\columnwidth]{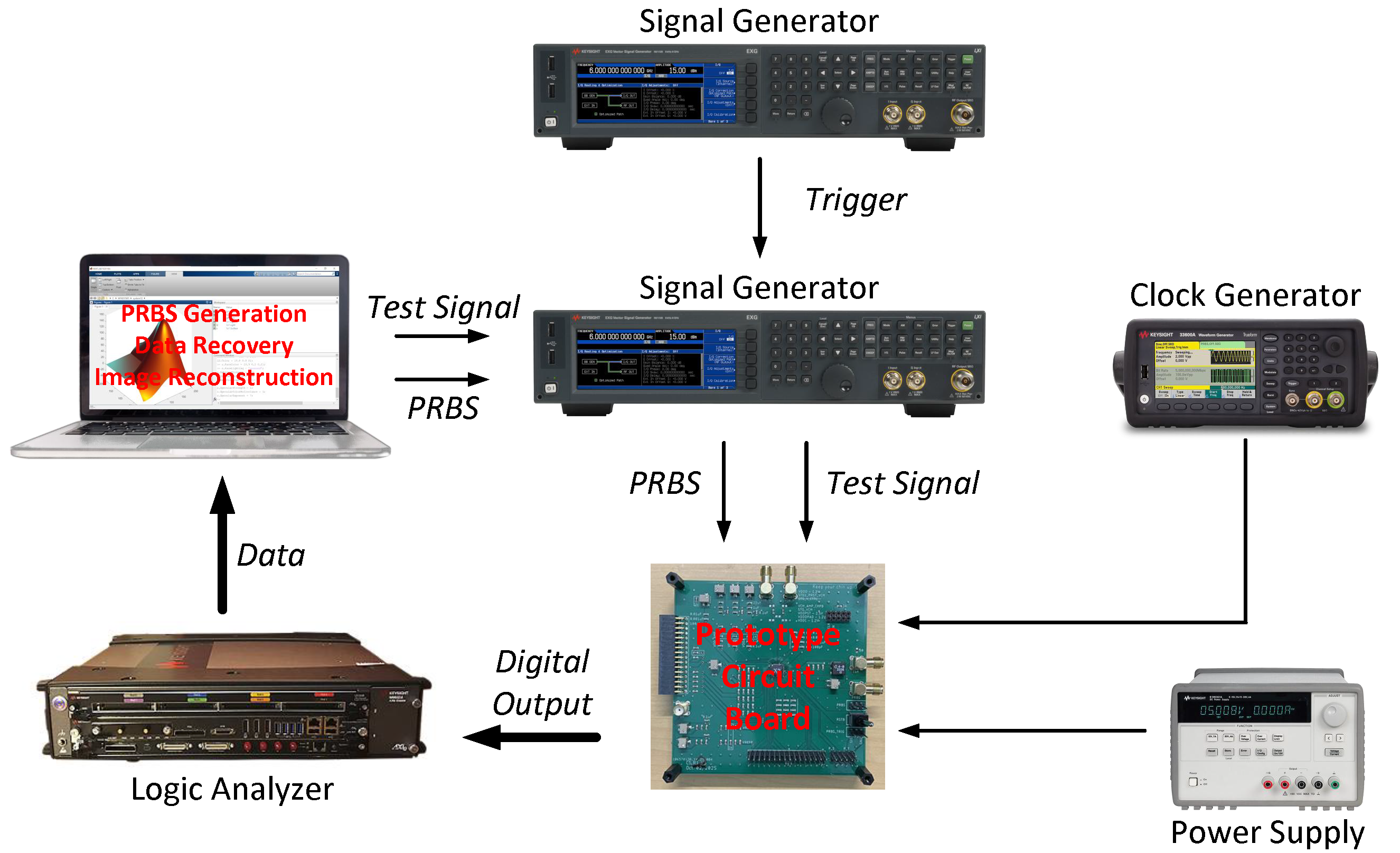}
\caption{End-to-end RF recovery and beamforming pipeline. FIELD II RF data are used as analog-equivalent input waveforms for the prototype CS-SAR ADC. The measured ADC outputs are recovered using the known PRBS sequence and pulse-dictionary model, and the reconstructed pre-beamformed RF data are then used for ultrasound beamforming.}
\label{fig:pipeline}
\end{figure}

\section{RF Recovery and Beamforming Pipeline}

The compressed ADC output is not a uniformly sampled RF trace when
$N_{cT}>1$. Each output code represents a PRBS-weighted
charge-domain combination of multiple neighboring RF samples. Therefore,
the measured ADC output is first recovered into an estimate of the
original pre-beamformed RF channel data, and the recovered RF data are
then passed to the same beamforming pipeline used for the non-compressed
reference.

For fair comparison across compression modes, the $N_{cT}=1$,
$N_{cT}=2$, and $N_{cT}=4$ datasets were processed using the
same RF recovery grid, interpolation method, beamforming geometry,
receive-aperture selection, envelope detection, log compression, and
image display settings. This ensures that the observed differences in
the final B-mode images are caused primarily by temporal compression and
RF recovery rather than by inconsistent downstream processing.

\subsection{Experimental Data Flow}

Fig.~\ref{fig:pipeline} summarizes the end-to-end recovery and imaging pipeline.
Ultrasound RF data were first generated using FIELD II and then used as
analog-equivalent input waveforms for the CS-SAR ADC experiment. The ADC
was operated in non-compressed and compressed modes using the
corresponding PRBS control sequence. The captured ADC outputs were then
recovered off chip using the known temporal sensing sequence and the
pulse-dictionary RF model described in Section II. Finally, the
recovered pre-beamformed RF channel data were beamformed to form B-mode
images.

The FIELD II RF data were generated at 250 MS/s. This high sampling rate
was used to provide a dense representation of the continuous-time RF
waveform applied to the ADC test setup. The nominal ADC sampling grid
was 10 MS/s, while the effective ADC output rates were 10 MS/s, 5 MS/s,
and 2.5 MS/s for $N_{cT}=1$, $N_{cT}=2$, and
$N_{cT}=4$, respectively. Thus, the 250-MS/s FIELD II sampling
rate should be interpreted as the waveform-generation reference, not as
the ADC output rate.

\subsection{FIELD II RF Data and Pulse Kernel}

RF data were generated using FIELD II \cite{jensen1992calculation}, \cite{jensen1996field}. The simulated
probe was a 64-element linear array with a 2.0-MHz center frequency and
50\% fractional bandwidth. Each element had a 5.0-mm elevation height,
0.3-mm pitch, and 0.03-mm kerf, corresponding to an active element width
of 0.27 mm. The wire-phantom dataset was generated using a single
plane-wave transmission, while the cyst-phantom dataset used 21 plane
waves steered from (\(- 20^{\circ}\)) to (\(+ 20^{\circ}\)).

The pulse kernel $h$ used in the RF dictionary was obtained from
the simulated array impulse response. This system-specific kernel was
used to construct the delayed-pulse dictionary described in Section II.
Using the probe-specific pulse response constrains the recovery to the
RF signal structure expected from the ultrasound acquisition chain,
rather than relying on a generic transform basis such as wavelets, DCT,
or wave atoms.

\begin{figure}[!t]
\centering
\includegraphics[width=0.88\columnwidth]{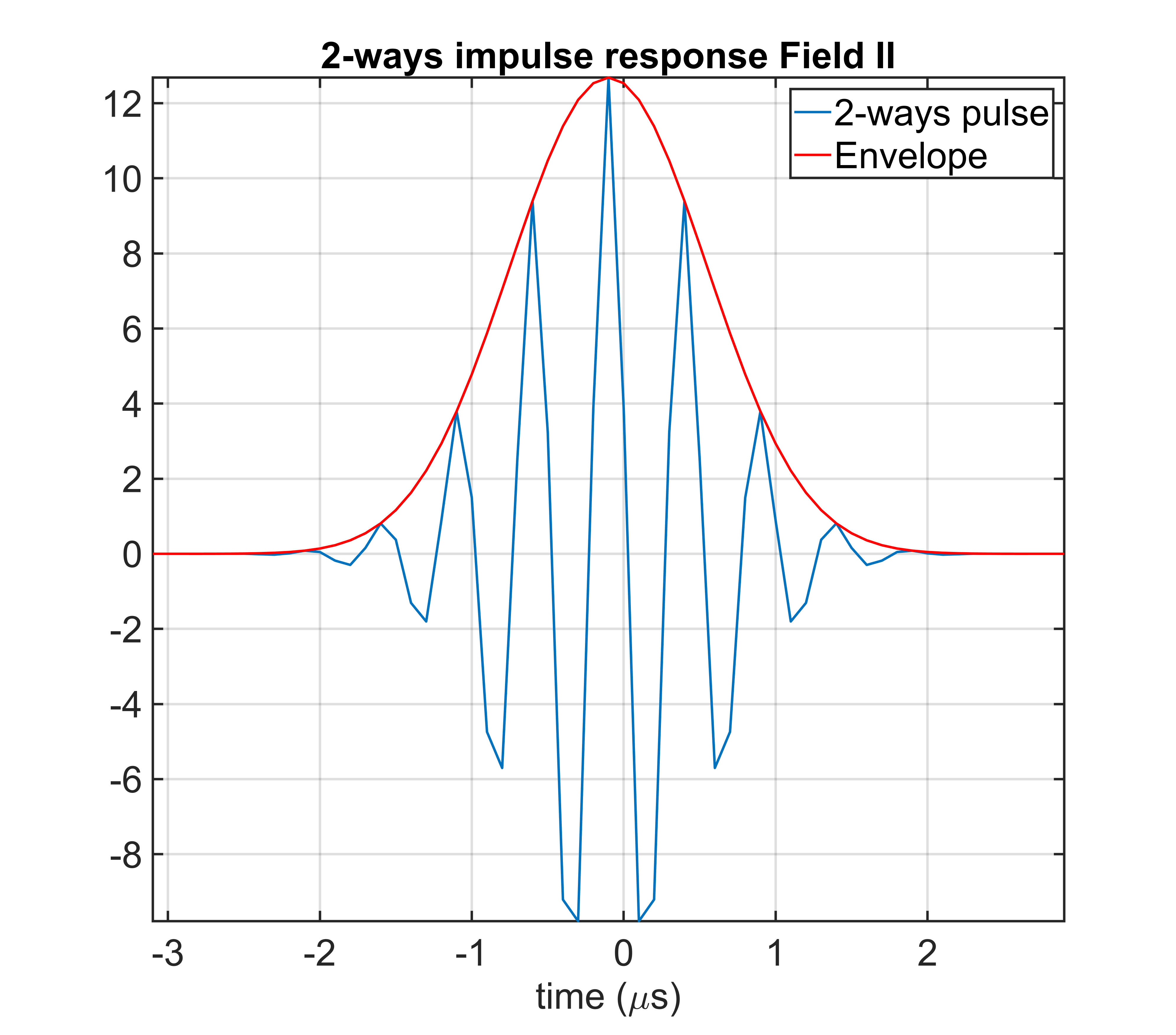}
\caption{Probe-specific pulse kernel used for RF recovery. The pulse response $h$ was obtained from the FIELD II array model and used to populate the delayed-pulse dictionary for pre-beamformed RF reconstruction.}
\label{fig:pulse_kernel}
\end{figure}

\subsection{RF Recovery From Measured ADC Outputs}

For each receive channel and transmit event, the measured ADC output was
recovered using the temporal sensing matrix determined by the applied
PRBS sequence and the compression ratio $N_{cT}$. The recovery
followed the pulse-dictionary elastic-net formulation described in
Section II. The same recovery procedure was used for all compressed
datasets, with $N_{cT}=1$ serving as the measured ASIC reference.

The regularization parameters were selected empirically and kept fixed
within each experiment. The selection was chosen to balance RF waveform
fidelity and suppression of reconstruction artifacts, without separately
optimizing the parameters for each displayed image. This choice avoids
overfitting the reconstruction to a particular target and provides a
consistent basis for comparing the $2\times$ and $4\times$ compression modes.

\begin{figure*}[!t]
\centering
\subfloat[Measured output spectrum.]{\includegraphics[width=0.43\textwidth]{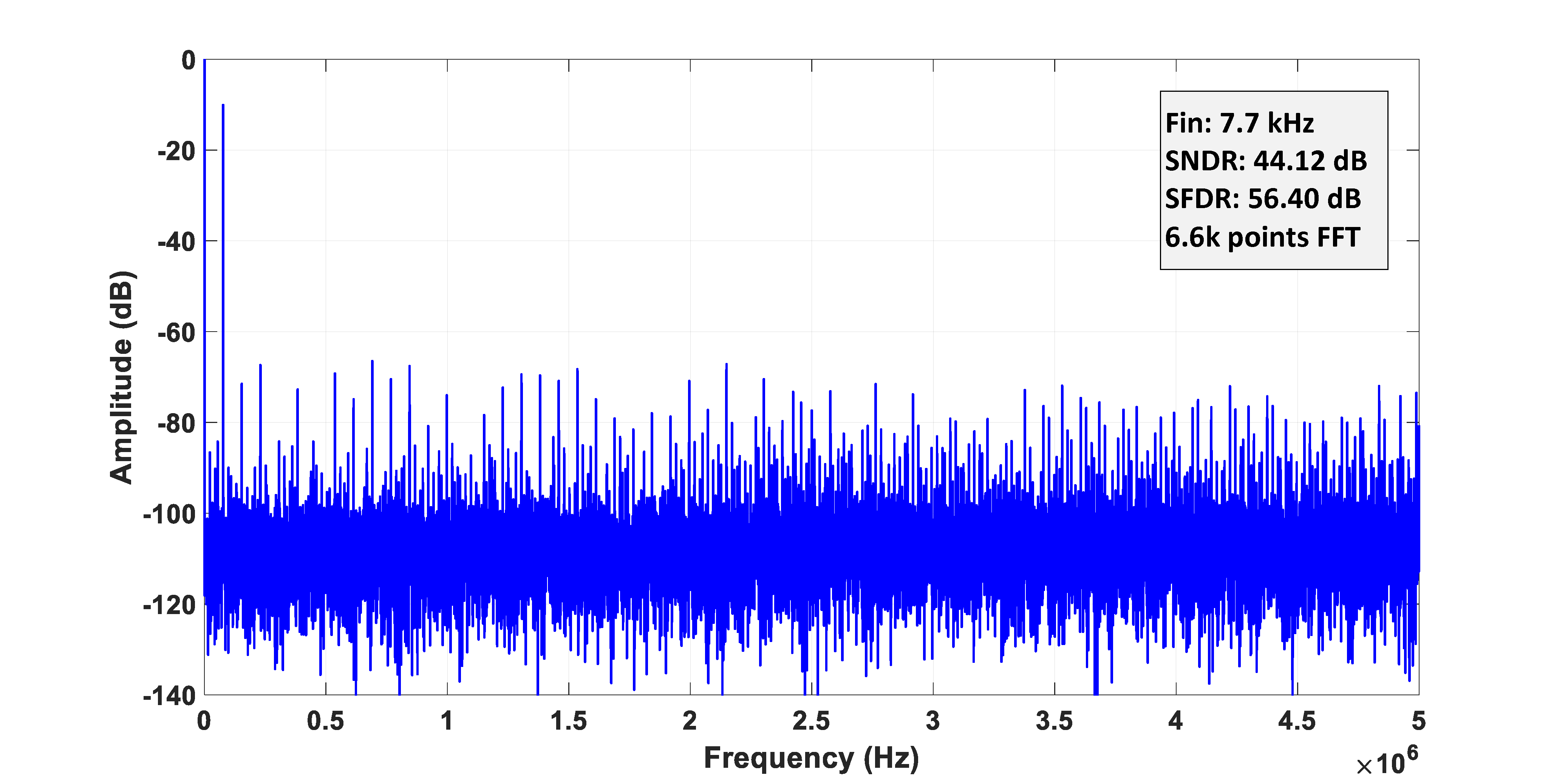}\label{fig:adc_spectrum}}
\hfil
\subfloat[SNDR versus input amplitude.]{\includegraphics[width=0.27\textwidth]{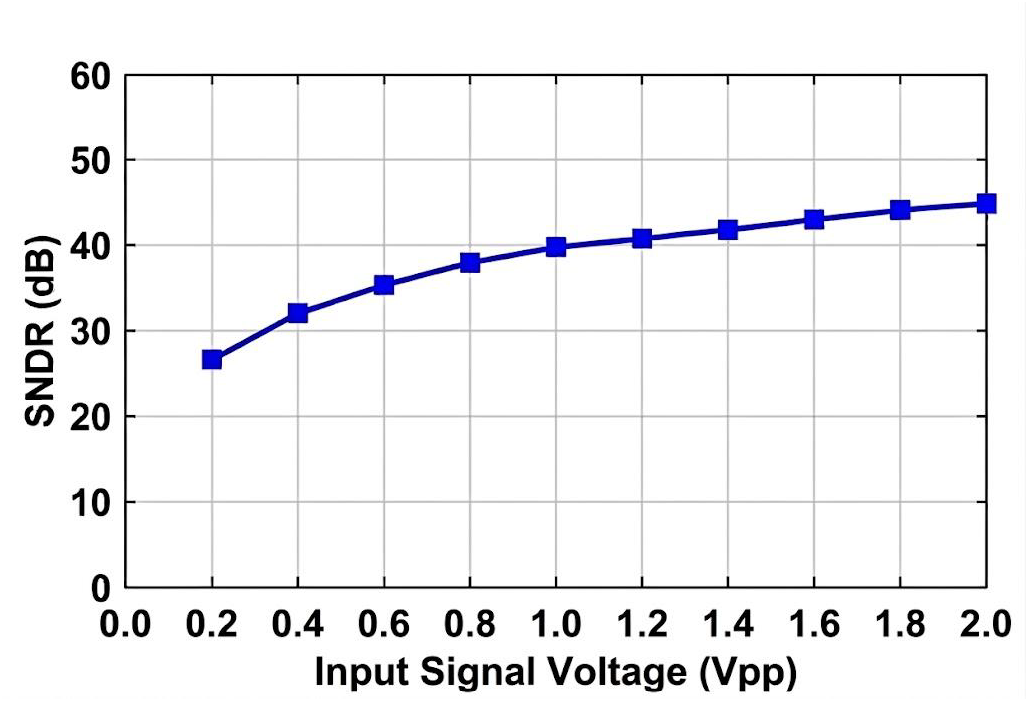}\label{fig:sndr_sweep}}
\hfil
\subfloat[Measured power breakdown.]{\includegraphics[width=0.25\textwidth]{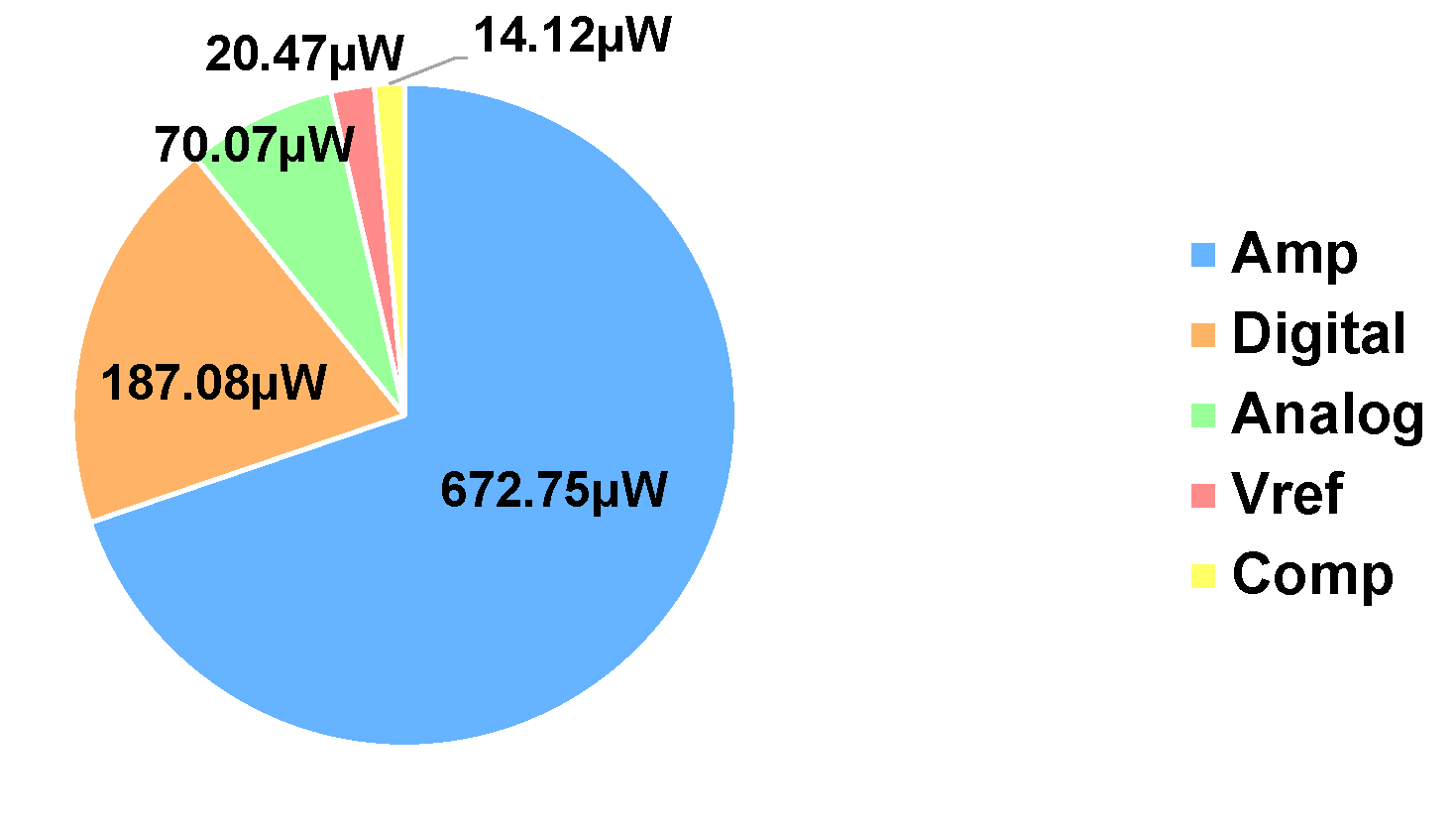}\label{fig:power_breakdown}}
\caption{Measured ADC characterization in the nominal non-compressed mode. The prototype was operated at 10 MS/s using a 50-MHz master clock, 1.2-V supply, and 1.2-V reference. The measured spectrum for a 7.7-kHz input yielded an SNDR of 44.12 dB and an SFDR of 56.40 dB.}
\label{fig:adc_characterization}
\end{figure*}

\subsection{Pixel-Based Plane-Wave Beamforming}

The recovered RF data were reconstructed into B-mode images using a
pixel-based plane-wave beamforming pipeline. For a pixel location
$p=(x,z)$ and a plane wave steered at angle (\(\theta_{k}\)), the
transmit delay was modeled as

\begin{equation}
\tau_{\mathrm{tx},k}(p)=\frac{x\sin\theta_k+z\cos\theta_k}{c}.
\label{eq:tx_delay}
\end{equation}

where $c$ is the assumed speed of sound. For receive element $i$
located at lateral position $x_i$, the receive delay was

\begin{equation}
\tau_{\mathrm{rx},i}(p)=\frac{\sqrt{(x-x_i)^2+z^2}}{c}.
\label{eq:rx_delay}
\end{equation}

The total sampling delay was therefore

\begin{equation}
\tau_{i,k}(p)=\tau_{\mathrm{tx},k}(p)+\tau_{\mathrm{rx},i}(p)+\tau_0.
\label{eq:total_delay}
\end{equation}

where \(\left( \tau_{0} \right)\) accounts for fixed transmit,
simulation, and acquisition-alignment offsets. The delayed RF sample was
obtained by interpolation from the recovered RF trace,

\begin{equation}
s_{i,k}(p)=\operatorname{interp}\!\left(\widehat{\mathbf{x}}_{i,k},\tau_{i,k}(p)\right).
\label{eq:interp}
\end{equation}

Dynamic receive-aperture selection was performed using an F-number of 1.
Therefore, only receiving elements within the selected aperture around
each image point were included in the beamforming sum. The same
receive-aperture rule was applied to all compression modes.

Delay-multiply-and-sum (DMAS) beamforming was used to combine delayed
receive-channel data \cite{matrone2015dmas}, \cite{matrone2016planewave}. For each transmit angle, the
DMAS output was computed from the delayed RF samples within the active
receive aperture \(\left( \mathcal{A}(p) \right)\) as

\begin{equation}
B_k(p)=\sum_{\substack{i,j\in\mathcal{A}(p)\\i<j}}
\operatorname{sgn}\!\left[s_{i,k}(p)s_{j,k}(p)\right]
\sqrt{\left|s_{i,k}(p)s_{j,k}(p)\right|}.
\label{eq:dmas}
\end{equation}

For the cyst-phantom dataset, the final compounded image was obtained by
coherently combining the beamformed outputs across all steering angles,

\begin{equation}
B(p)=\sum_{k=1}^{N_{\theta}}B_k(p).
\label{eq:compound}
\end{equation}

where (\(N_{\theta} = 21\)) for the cyst-phantom experiment. The
compounded RF image was then envelope-detected, log-compressed, and
displayed using a fixed dynamic range. Axial Wiener filtering was
applied as a fixed post-processing step to all compared images.

\begin{figure*}[!t]
\centering
\subfloat[$N_{cT}=2$.]{\includegraphics[width=0.47\textwidth]{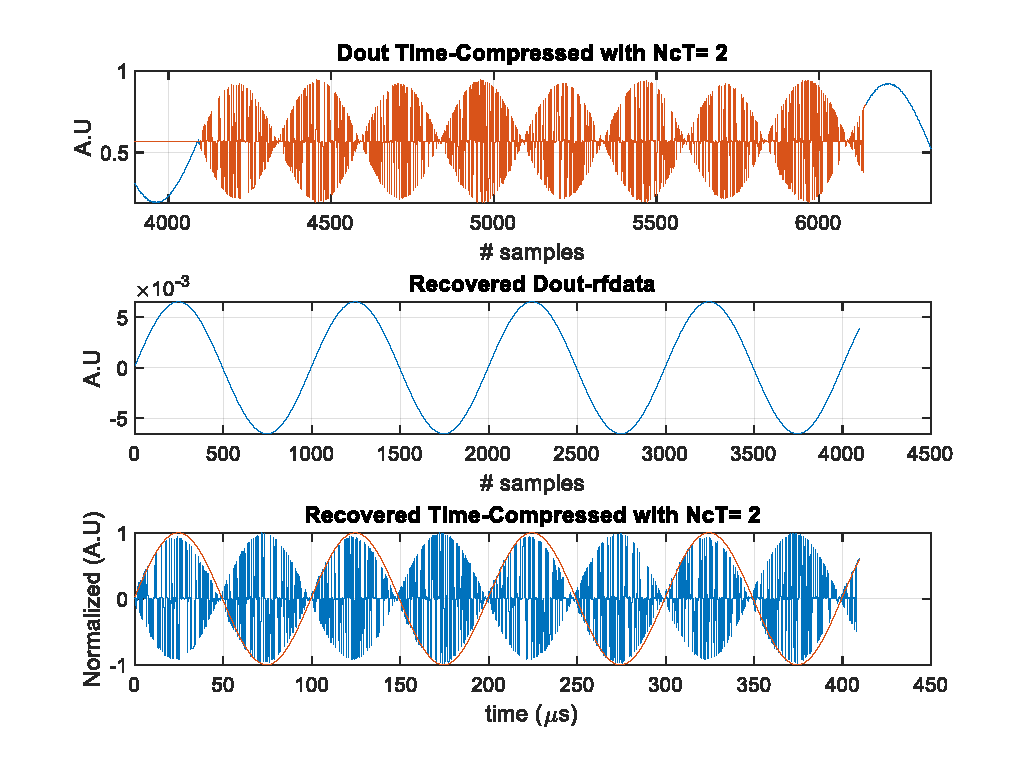}\label{fig:sin_nct2}}
\hfil
\subfloat[$N_{cT}=4$.]{\includegraphics[width=0.47\textwidth]{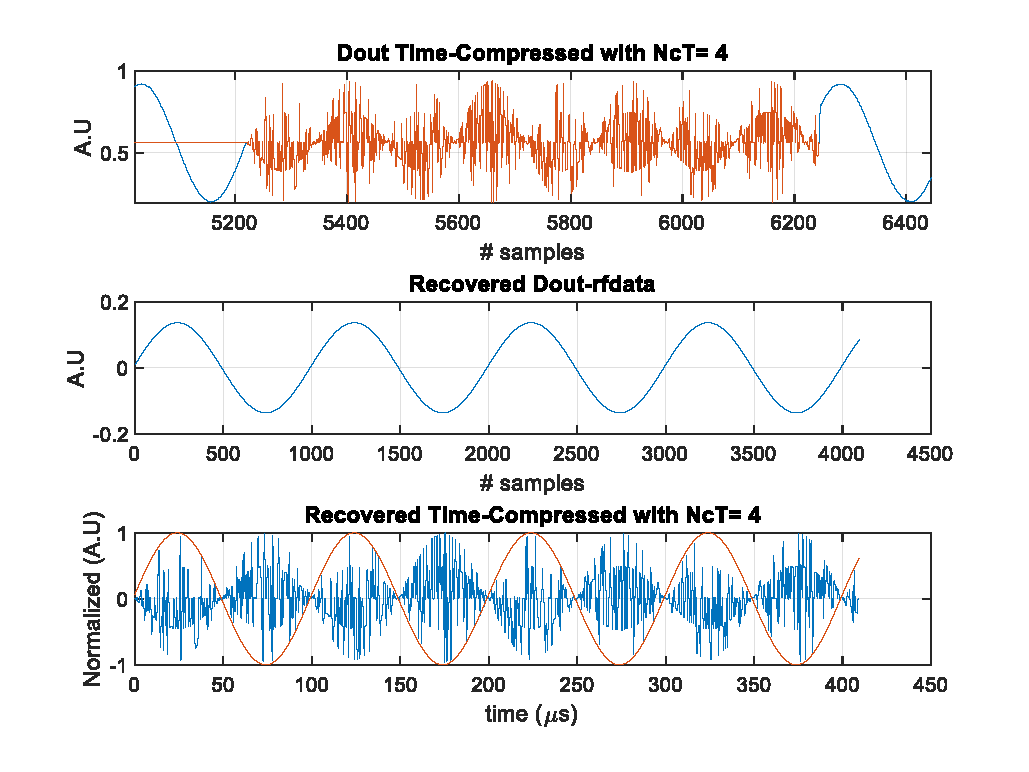}\label{fig:sin_nct4}}
\caption{Sinusoidal waveform recovery using measured ADC outputs. A 10-kHz input waveform was acquired in compressed modes and reconstructed using the known PRBS sequence. The recovered waveforms verify the temporal compression and recovery chain.}
\label{fig:sin_recovery}
\end{figure*}

\subsection{Quantitative Evaluation}

The quantitative analysis was divided into RF-domain and image-domain
metrics. RF-domain metrics evaluate whether temporal compression and
recovery preserve the waveform information required for beamforming.
Image-domain metrics evaluate whether the recovered RF data preserve
point-target geometry, spatial resolution, low-contrast lesion
separability, and similarity to the non-compressed reference image.

For RF analysis, normalized root-mean-square error (NRMSE) and normalized
cross-correlation (NCC) were computed for all available RF traces rather
than for a selected best-performing trace. For each compression mode,
the resulting trace-wise values were summarized by the median and the
first and third quartiles [Q1, Q3]. The two metrics quantify relative
waveform error and mean-removed waveform-shape similarity, respectively:

\begin{equation}
\mathrm{NRMSE}=\frac{\|\widehat{\mathbf{x}}-\mathbf{x}_{\mathrm{ref}}\|_2}
{\|\mathbf{x}_{\mathrm{ref}}\|_2}.
\label{eq:nrmse}
\end{equation}

\begin{equation}
\mathrm{NCC}=\frac{(\mathbf{x}_{\mathrm{ref}}-\overline{\mathbf{x}}_{\mathrm{ref}})^T
(\widehat{\mathbf{x}}-\overline{\widehat{\mathbf{x}}})}
{\|\mathbf{x}_{\mathrm{ref}}-\overline{\mathbf{x}}_{\mathrm{ref}}\|_2
\|\widehat{\mathbf{x}}-\overline{\widehat{\mathbf{x}}}\|_2}.
\label{eq:ncc}
\end{equation}

In \eqref{eq:nrmse} and \eqref{eq:ncc}, $\mathbf{x}_{\mathrm{ref}}$
denotes the reference RF waveform obtained from the $N_{cT}=1$
acquisition, and $\widehat{\mathbf{x}}$ denotes the RF waveform recovered
from the compressed ADC measurements. The operator $\|\cdot\|_2$
represents the Euclidean norm, while
$\overline{\mathbf{x}}_{\mathrm{ref}}$ and
$\overline{\widehat{\mathbf{x}}}$ denote the corresponding mean values.
An ideal recovery gives $\mathrm{NRMSE}=0$ and $\mathrm{NCC}=1$.

For the wire phantom, axial and lateral full-width at half-maximum
(FWHM) were measured from the same wire target in all compression modes.
Localization error was evaluated using the position of this wire in the
$N_{cT}=1$ reference,

\begin{equation}
\mathbf{r}_{\mathrm{ref}}=(x_{\mathrm{ref}},z_{\mathrm{ref}}),
\end{equation}

and in the compressed reconstruction,

\begin{equation}
\mathbf{r}_{\mathrm{CS}}=(x_{\mathrm{CS}},z_{\mathrm{CS}}).
\end{equation}

The signed lateral and axial displacements were

\begin{equation}
\Delta x=x_{\mathrm{CS}}-x_{\mathrm{ref}},\qquad
\Delta z=z_{\mathrm{CS}}-z_{\mathrm{ref}},
\end{equation}

and the scalar localization error was computed as

\begin{equation}
e_{\mathrm{loc}}=\sqrt{(\Delta x)^2+(\Delta z)^2},
\label{eq:localization_error}
\end{equation}

with all spatial quantities reported in millimeters.

For the cyst phantom, contrast-to-noise ratio (CNR), generalized CNR
(gCNR), and structural similarity index (SSIM) relative to the
$N_{cT}=1$ reference were used to quantify low-contrast and overall image
fidelity. For regions of interest selected inside the cyst and in the
surrounding speckle background,

\begin{equation}
\mathrm{CNR}=\frac{|\mu_{\mathrm{bg}}-\mu_{\mathrm{cyst}}|}
{\sqrt{\sigma_{\mathrm{bg}}^2+\sigma_{\mathrm{cyst}}^2}}.
\label{eq:cnr}
\end{equation}

where $\mu_{\mathrm{bg}}$ and $\mu_{\mathrm{cyst}}$ are the mean image
amplitudes in the background and cyst regions, and $\sigma_{\mathrm{bg}}$
and $\sigma_{\mathrm{cyst}}$ are the corresponding standard deviations.
The gCNR was computed from the overlap between the normalized cyst and
background intensity histograms,

\begin{equation}
\mathrm{gCNR}=1-\sum_q\min\!\left[p_{\mathrm{cyst}}(q),p_{\mathrm{bg}}(q)\right].
\label{eq:gcnr}
\end{equation}

where $p_{\mathrm{cyst}}$ and $p_{\mathrm{bg}}$ are normalized histogram
probabilities over intensity bin $q$. SSIM provides a complementary
measure of overall structural similarity to the $N_{cT}=1$ image. The
same beamforming, post-processing, normalization, and image grid were
used for all compression modes.

\section{Results}

The prototype was evaluated in the following manner. First, the ADC was
characterized electrically in the non-compressed mode to establish the
measured converter baseline. Second, sinusoidal waveform recovery was
used to verify the temporal compression and reconstruction chain
independent of ultrasound propagation. Third, FIELD-II-generated RF data
from a wire phantom were used to evaluate RF-line recovery and timing
fidelity. Finally, recovered RF data were beamformed to form B-mode
images from wire and cyst phantoms. The wire phantom was used to assess
point-target localization and resolution, while the cyst phantom was
used as the primary imaging validation because it contains anechoic
structures embedded in a speckle-rich background.
\begin{table}[!t]
\caption{FIELD II, Recovery, and Beamforming Parameters}
\label{tab:imaging_parameters}
\centering
\footnotesize
\begin{tabular}{@{}ll@{}}
\toprule
Parameter & Value\\
\midrule
Center frequency & 2.0 MHz\\
Fractional bandwidth & 50\%\\
Number of elements & 64\\
Element pitch & 0.3 mm\\
Kerf & 0.03 mm\\
Element height & 5.0 mm\\
FIELD II sampling rate & 250 MS/s\\
ADC nominal sampling grid & 10 MS/s\\
Compression ratios & $N_{cT}=1,2,4$\\
Wire-phantom transmit sequence & Single plane wave\\
Cyst-phantom transmit sequence & 21 plane waves, $-20^{\circ}$ to $+20^{\circ}$\\
Receive F-number & 1\\
Beamformer & DMAS with coherent compounding\\
\bottomrule
\end{tabular}
\end{table}

\subsection{ADC Electrical Characterization}

The fabricated CS-SAR ADC was first measured in the nominal
non-compressed mode, $N_{cT}=1$. The prototype was operated with a
50-MHz master clock, 1.2-V supply, and 1.2-V reference. In this mode,
the ADC produced a 10-MS/s output rate and consumed 964.49
(\(\mu\text{W}\)). The inter-stage residue amplifier was the dominant
power-consuming block, followed by the digital control circuitry,
remaining analog circuits, reference buffer, and comparators.

For a 7.7-kHz sinusoidal input, the measured output spectrum yielded an
SNDR of 44.12 dB and an SFDR of 56.40 dB. These results define the
converter-level baseline used for the subsequent temporal compression
and RF recovery experiments. The measured SNDR limitation is discussed
in Section VI in relation to charge-redistribution timing and
residue-generation nonidealities.

\begin{figure*}[!t]
\centering
\subfloat[$N_{cT}=2$.]{\includegraphics[width=0.47\textwidth]{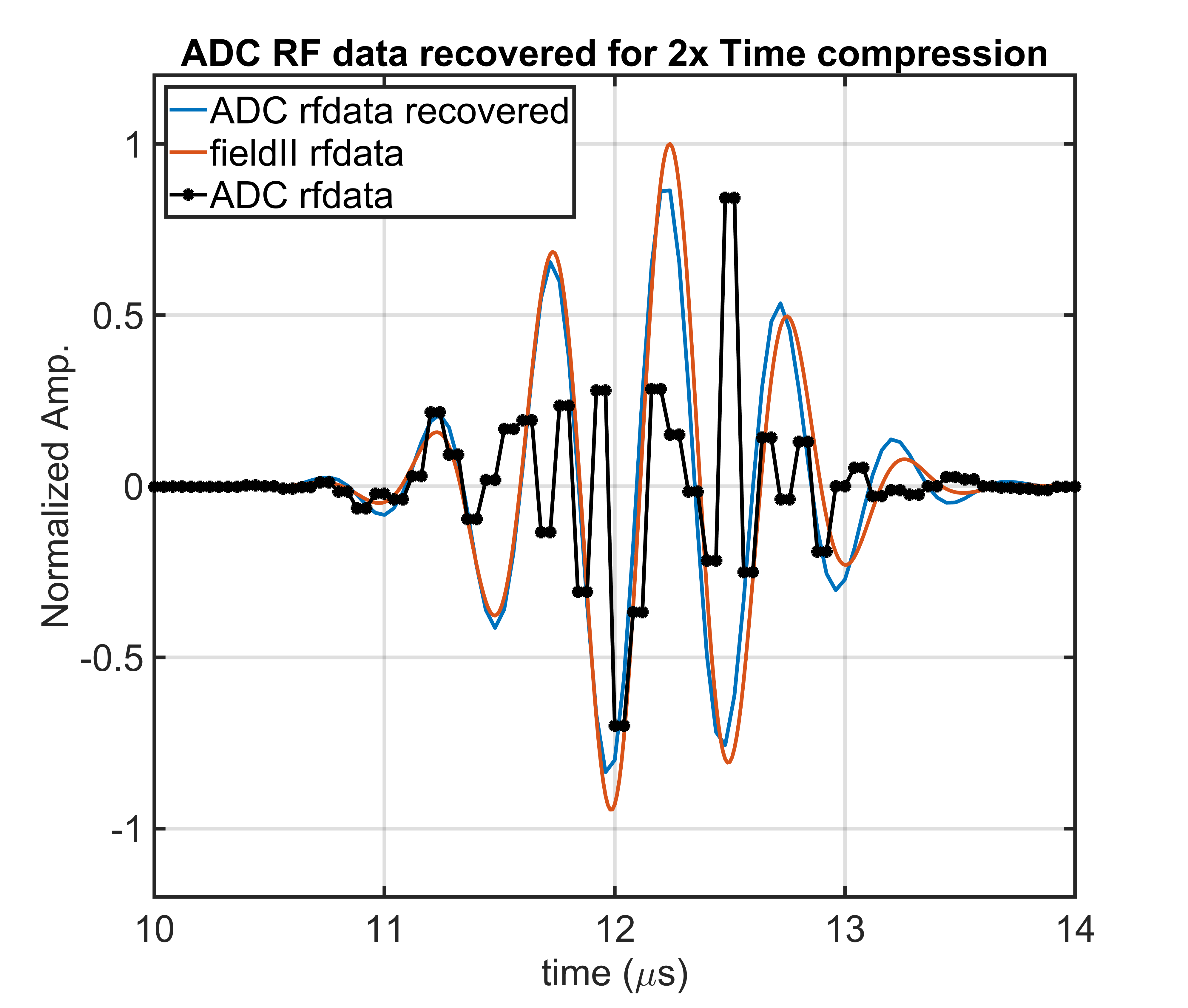}\label{fig:rf_nct2}}
\hfil
\subfloat[$N_{cT}=4$.]{\includegraphics[width=0.47\textwidth]{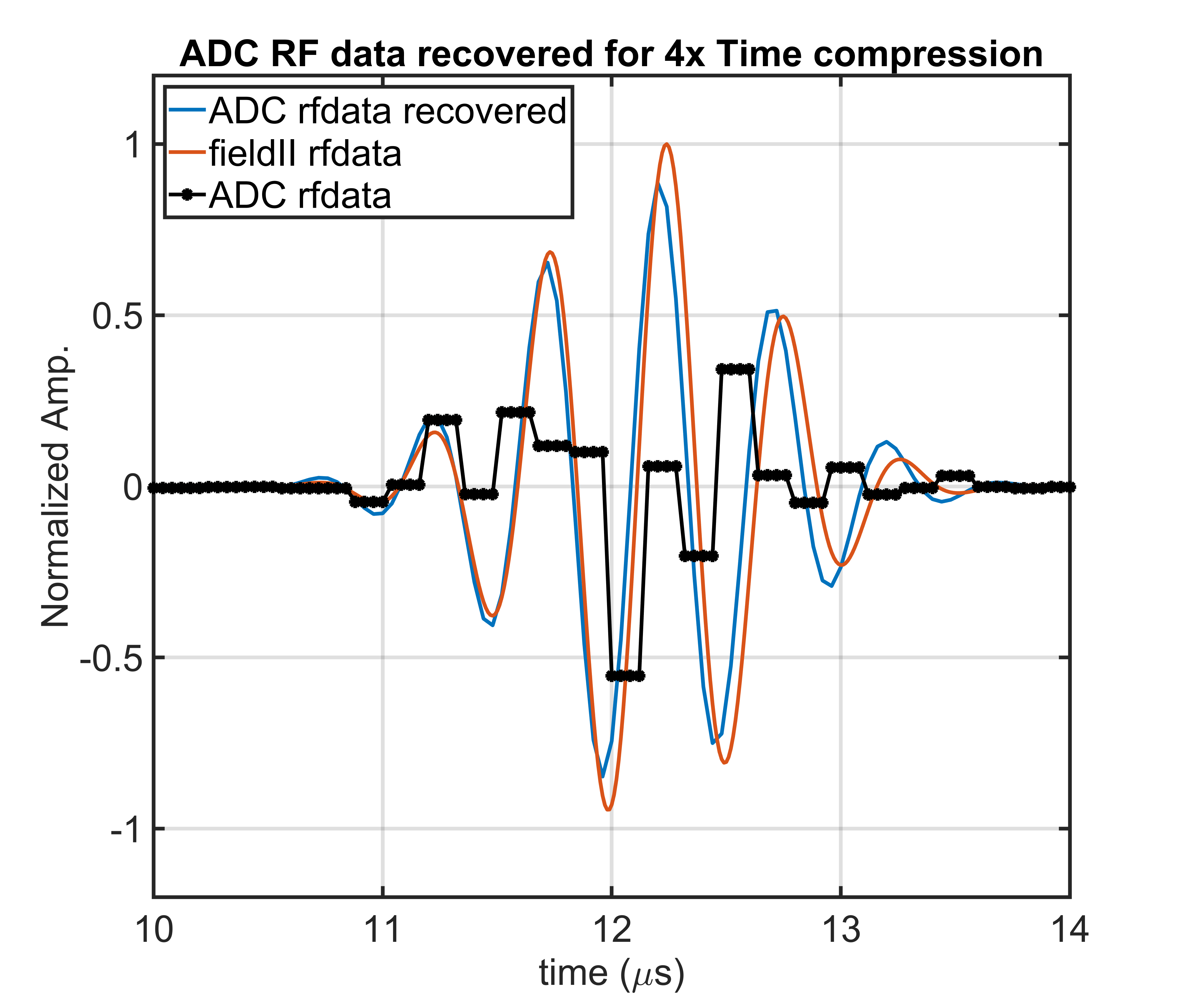}\label{fig:rf_nct4}}
\caption{RF-line recovery from FIELD-II-generated wire-phantom data using measured ADC outputs. A representative middle-array channel is shown for the reference RF trace and the recovered compressed traces.}
\label{fig:rf_recovery}
\end{figure*}

\subsection{Sinusoidal Temporal CS Recovery}

A 10-kHz sinusoidal input was used to verify the temporal compression
and recovery chain. This experiment isolates the ADC compression
sequence, PRBS control, digital output capture, and off-chip
reconstruction from ultrasound propagation and beamforming effects.

The raw compressed ADC outputs represent coded PRBS-weighted
measurements rather than direct samples of the input sinusoid. After
recovery using the known temporal sensing sequence, the reconstructed
waveforms followed the original sinusoidal input for both compression
factors. This confirms that the physical CS-SAR sampling operation and
the digital recovery model are mutually consistent before applying the
method to ultrasound RF data.

\subsection{RF-Line Recovery From Wire-Phantom Data}

The next validation step used FIELD-II-generated ultrasound RF data from
the 64-element linear array described in Section IV. The RF waveform was
generated at 250 MS/s and applied to the ADC as an analog-equivalent
input. A single plane-wave transmission was used for the wire-phantom
RF-line experiment to provide a controlled pulse-echo signal for timing
and waveform-fidelity evaluation.

For each compression mode, the same PRBS sequence used during ADC
acquisition was used in the recovery matrix. A middle-array receive
channel is shown in Fig.~\ref{fig:rf_recovery} only as a representative
visual example of RF waveform recovery. The quantitative NRMSE and NCC
results, however, were computed over all available RF traces and are
reported as distribution summaries in Table~\ref{tab:metrics}; no
best-performing RF trace was selected for the quantitative comparison.

For $N_{cT}=2$, the representative recovered RF signal preserves the
main pulse timing and waveform structure of the reference trace. At
$N_{cT}=4$, the principal echo remains recoverable, although residual
waveform error is more apparent. The population-level RF metrics reported
below quantify this compression-dependent loss of waveform fidelity.

\subsection{Wire-Phantom B-Mode Reconstruction}

Recovered RF data from all receive elements were beamformed to form
wire-phantom B-mode images for $N_{cT}=1$, $N_{cT}=2$, and
$N_{cT}=4$. The $N_{cT}=1$ image served as the measured
no-compression ASIC reference. The wire phantom is a favorable target
for CS because it is sparse and high contrast; therefore, this
experiment is used mainly to evaluate point-target localization,
resolution preservation, and artifact behavior.

The $N_{cT}=2$ reconstruction preserved the wire locations and
overall image structure relative to the no-compression reference. The
$N_{cT}=4$ reconstruction also retained the principal wire locations,
although background artifacts around weaker reflectors became more
apparent. As quantified below, the measured FWHM and localization error
remain stable across compression modes, indicating that the principal
penalty is not appreciable point-target broadening or displacement.

\begin{figure*}[!t]
\centering
\includegraphics[width=0.98\textwidth]{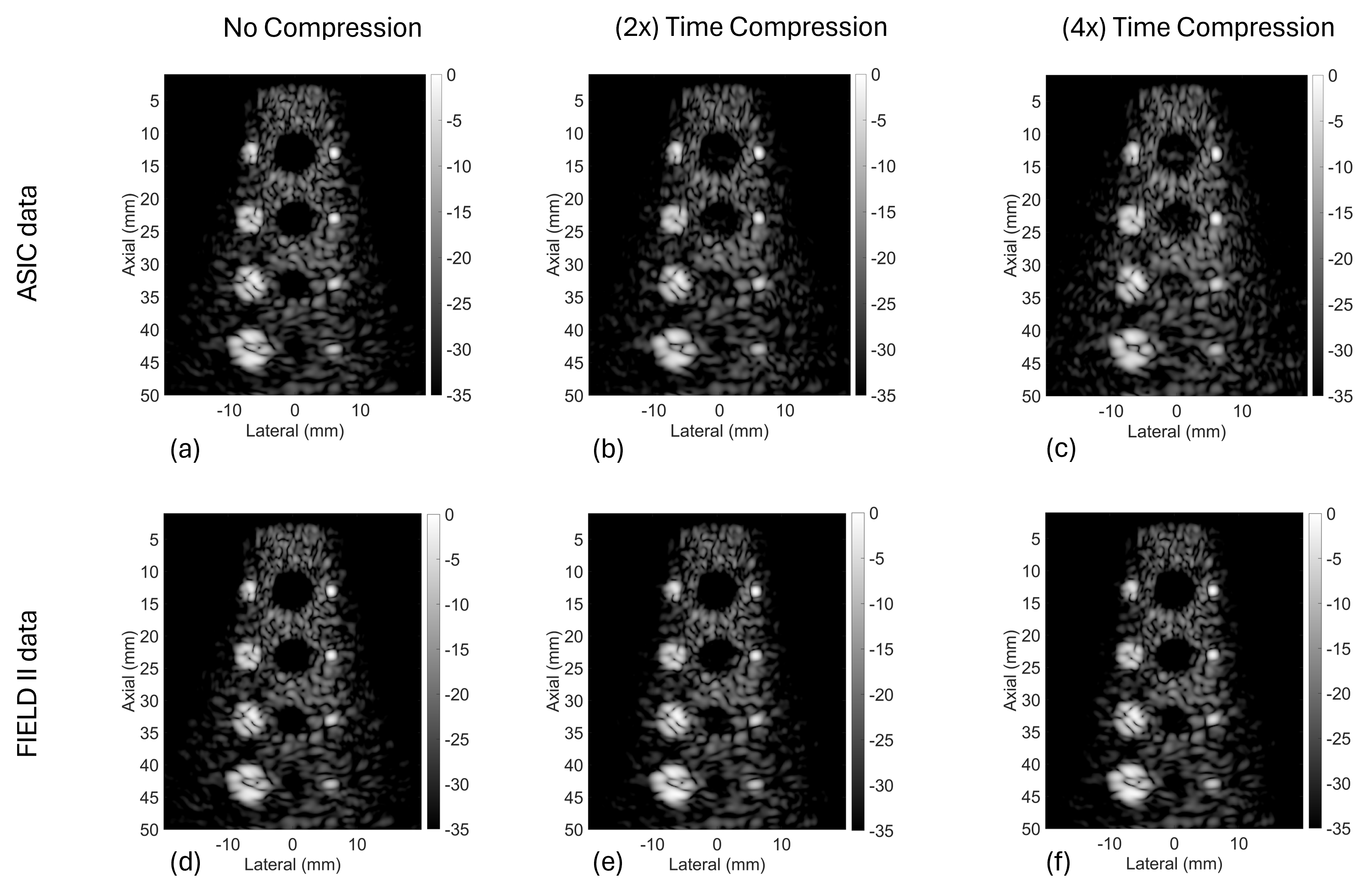}
\caption{Cyst-phantom B-mode reconstructions using identical beamforming, post-processing, normalization, and display dynamic range. The top row shows images reconstructed from measured ASIC outputs and the bottom row shows the corresponding FIELD II data: no compression, $N_{cT}=1$ (a,d); $2\times$ temporal compression, $N_{cT}=2$ (b,e); and $4\times$ temporal compression, $N_{cT}=4$ (c,f). The compressed ASIC modes reduce the ADC output rates from the nominal 10 MS/s to 5 MS/s and 2.5 MS/s, respectively.}
\label{fig:cyst_bmode}
\end{figure*}

\subsection{Cyst-Phantom B-Mode Reconstruction}

The cyst phantom provides the main imaging validation of the proposed
temporal CS acquisition approach. Unlike wire targets, the cyst phantom
contains anechoic or low-echo regions embedded in a speckle-rich
background. It therefore tests whether the recovered RF data preserve
lesion visibility, boundary definition, and speckle structure after
compressed acquisition and beamforming.

For this experiment, RF data were generated using 21 plane-wave
transmissions steered from \(\left( - 20^{\circ} \right)\) to
\(\left( + 20^{\circ} \right)\). The recovered RF data were
reconstructed using the pixel-based DMAS compounding pipeline described
in Section IV, with the same F-number, post-processing, normalization,
and display settings applied to all compression modes.

The no-compression $N_{cT}=1$ reconstruction provides the measured
ASIC reference. With $N_{cT}=2$, the cyst structures remain visible and
the speckle background is largely preserved, although local differences
in texture and contrast are observed relative to the reference. With
$N_{cT}=4$, the main cyst regions remain identifiable, demonstrating
that fourfold temporal compression can still support B-mode image
reconstruction. However, the stronger background granularity and residual
artifacts are accompanied by the larger CNR, gCNR, and SSIM degradation
reported in Table~\ref{tab:metrics}.

\subsection{Quantitative RF and Image Metrics}

Table~\ref{tab:metrics} summarizes the RF- and image-domain metrics across
the three temporal compression modes. Because the RF statistics include
all available RF traces, the reported medians and quartiles characterize
the full evaluated dataset rather than a favorable example. At
$N_{cT}=2$, the median RF NRMSE is 0.36 [0.33, 0.61] and the median NCC
is 0.981 [0.920, 0.988]. At $N_{cT}=4$, the NRMSE increases to 0.56
[0.48, 0.62] while the NCC decreases to 0.932 [0.893, 0.967], showing a
progressive loss of RF waveform fidelity with increasing compression.

Despite this RF-domain degradation, the point-target measurements remain
remarkably stable. The axial FWHM changes from 1.074 mm in the reference
to 1.058 mm and 1.028 mm for $N_{cT}=2$ and $N_{cT}=4$, respectively,
while the lateral FWHM changes from 1.117 mm to 1.107 mm and 1.123 mm.
These few-percent differences should not be interpreted as improved
resolution with compression; rather, no appreciable degradation in axial
or lateral FWHM is observed. The corresponding localization errors are
only 0.032 mm and 0.036 mm, confirming that point-target position is
preserved even in the more aggressive compression mode.

The cyst-phantom metrics are more sensitive to temporal compression. CNR
decreases from 3.534 in the reference to 2.047 at $N_{cT}=2$ and 1.379
at $N_{cT}=4$, corresponding to retention of 57.9\% and 39.0\% of the
reference CNR, respectively. The gCNR decreases from 0.980 to 0.837 and
0.702, retaining 85.4\% and 71.6\% of the reference value. SSIM follows
an intermediate trend, decreasing from 1 to 0.94 and 0.87. Thus, the
measured data show a clear separation between robust point-target
geometry/resolution and the greater compression sensitivity of
low-contrast cyst separability and broader image structure.

\begin{table}[!t]
\caption{Quantitative RF and Image-Quality Metrics for Compressed Acquisition}
\label{tab:metrics}
\centering
\footnotesize
\setlength{\tabcolsep}{2.5pt}
\renewcommand{\arraystretch}{1.10}

\begin{tabularx}{0.98\columnwidth}{@{}
>{\raggedright\arraybackslash}X
*{3}{>{\centering\arraybackslash}p{0.205\columnwidth}}
@{}}
\toprule
Metric & $N_{cT}=1$ & $N_{cT}=2$ & $N_{cT}=4$ \\
\midrule

RF NRMSE, median [Q1, Q3]
& 0
& \shortstack{0.36\\[-1pt]{\scriptsize [0.33, 0.61]}}
& \shortstack{0.56\\[-1pt]{\scriptsize [0.48, 0.62]}} \\

RF NCC, median [Q1, Q3]
& 1
& \shortstack{0.981\\[-1pt]{\scriptsize [0.92, 0.988]}}
& \shortstack{0.932\\[-1pt]{\scriptsize [0.893, 0.967]}} \\

Axial FWHM (mm)
& 1.074 & 1.058 & 1.028 \\

Lateral FWHM (mm)
& 1.117 & 1.107 & 1.123 \\

Localization error (mm)
& 0 & 0.032 & 0.036 \\

Cyst CNR
& 3.534 & 2.047 & 1.379 \\

Cyst gCNR
& 0.980 & 0.837 & 0.702 \\

SSIM
& 1 & 0.94 & 0.87 \\

\bottomrule
\end{tabularx}
\end{table}

\section{Discussion}

The proposed CS-SAR ADC demonstrates a compressed-sensing medical
ultrasound receiver based on SAR ADC architecture that performs temporal
compression and sub-Nyquist sampling during RF acquisition. The central
contribution is not post-acquisition digital compression, but
acquisition-side coding of consecutive RF samples before SAR conversion.
By embedding PRBS-controlled polarity selection and charge-domain
accumulation into the ADC sampling network, the prototype reduces the
number of SAR conversion events and output samples by the temporal
compression ratio \(\left( N_{cT} \right)\). The recovered
pre-beamformed RF data can then be processed using conventional
ultrasound beamforming pipelines.

This work differs from prior ultrasound CS studies that primarily
demonstrated algorithmic sub-sampling, compressed beamforming, or
off-line reconstruction using conventionally acquired RF data. It also
differs from analog downconversion or subarray beamforming approaches in
which the sampled bandwidth is reduced after frequency translation or
aperture-domain preprocessing. In the present work, the ADC itself
generates coded temporal measurements of the RF waveform before
digitization. In addition, unlike sparse-target demonstrations that rely
mainly on wires, point reflectors, or hair-like absorbers, the
cyst-phantom experiment evaluates whether the recovered RF data can
preserve image structure in a speckle-rich, non-point-target scene.

To the best of our knowledge, this is the first measured on-chip
SAR-ADC-based temporal compressive acquisition demonstration for
pulse-echo medical ultrasound RF imaging that recovers pre-beamformed RF
data and validates B-mode reconstruction using both sparse wire targets
and a speckle-rich cyst phantom. The claim is intentionally limited to
temporal CS acquisition; spatial compression is not demonstrated in the
present silicon.

The wire-phantom results provide a controlled validation of timing
fidelity, point-target localization, and resolution preservation.
Because wires are sparse and high contrast, they are useful for
evaluating whether temporal compression introduces pulse shifts,
broadening, or artifacts, but they are not sufficient to establish
performance in realistic ultrasound scenes. Therefore, the wire results
should be interpreted as a functionality and point-spread-function
validation.

The combined RF and image metrics reveal that temporal compression does
not affect all forms of fidelity equally. At $N_{cT}=2$, the median NCC
remains high at 0.981 even though the median NRMSE is 0.36. These values
are not contradictory: NRMSE is sensitive to absolute amplitude and
sample-wise reconstruction errors, whereas NCC emphasizes mean-removed
waveform shape and timing similarity. The high NCC is therefore
consistent with the small 0.032-mm localization error and the essentially
unchanged axial and lateral FWHM, even though the recovered RF waveform
is not sample-by-sample identical to the reference.

At $N_{cT}=4$, both RF metrics show stronger degradation, with a median
NRMSE of 0.56 and median NCC of 0.932. Nevertheless, the localization
error remains only 0.036 mm and the FWHM values remain within a few
percent of the reference. In contrast, the cyst metrics deteriorate more
substantially: CNR is reduced by 61.0\% and gCNR by 28.4\% relative to
the $N_{cT}=1$ reference, while SSIM decreases to 0.87. The results
therefore distinguish geometric fidelity from contrast/statistical
fidelity. The compressed measurements retain sufficient waveform timing
and structure to localize a strong reflector and preserve nominal
point-target resolution, while reconstruction errors increasingly alter
the weaker RF components and intensity distributions that determine
speckle statistics and low-contrast cyst conspicuity.

This distinction also clarifies the value of the cyst-phantom experiment.
A wire-only validation would suggest little image-domain penalty because
FWHM and localization remain nearly unchanged even at $N_{cT}=4$. The
speckle-rich cyst phantom reveals a degradation mode that point targets
cannot expose. In this sense, $N_{cT}=2$ provides the more favorable
high-fidelity operating point, combining strong RF correlation, preserved
point-target geometry, and SSIM of 0.94, whereas $N_{cT}=4$ represents a
more aggressive data-reduction mode in which geometric information
remains robust but low-contrast lesion separability is more clearly
compromised.

The RF quartiles provide an additional indication of reconstruction
variability. For $N_{cT}=2$, the NRMSE median is 0.36 while Q3 reaches
0.61, indicating that a portion of the evaluated traces has substantially
larger normalized error than the median trace. One possible contributor
is variation in RF trace energy, because normalization by
$\|\mathbf{x}_{\mathrm{ref}}\|_2$ makes NRMSE more sensitive for weak
reference signals. This mechanism was not isolated in the present study
and should therefore be treated as a hypothesis rather than a measured
cause. Future work should examine noise- and signal-energy-aware
selection of the elastic-net parameters $\lambda$ and $\alpha$, including
potential angle-dependent tuning for compounded plane-wave acquisitions.

A key advantage of the proposed architecture is that the number of SAR
conversions decreases with \(\left( N_{cT} \right)\). For a fixed RF
acquisition window, the number of ADC output codes scales approximately
as \(\left( 1/N_{cT} \right)\). Therefore, the temporal compression
modes directly reduce ADC output data volume from 10 MS/s in the
non-compressed mode to 5 MS/s and 2.5 MS/s in the $2\times$ and $4\times$ modes,
respectively. However, the measured total power of the present prototype
may not scale ideally with \(\left( N_{cT} \right)\). The inter-stage
residue amplifier is the dominant power-consuming block and remains
active in the current implementation. As a result, the prototype
demonstrates reduced conversion count and output data rate, but it does
not yet fully realize the theoretical power scaling that would be
possible if all conversion-related and residue-processing blocks were
duty-cycled or power-gated in compression mode. A future optimized
implementation could use amplifier duty cycling, compressed-mode residue
scheduling, clock gating, and stronger power-domain partitioning to
translate temporal compression more directly into total energy savings.

The charge-domain implementation does not necessarily impose a
first-order $kT/C$ noise penalty when the sampling capacitance is
partitioned across the compression window. If \(\left( N_{cT} \right)\)
independent samples are acquired onto capacitance \((C_{tot}/N_{cT}\))
and then averaged by charge redistribution, the equivalent noise
variance after averaging can be approximated as

\begin{equation}
\sigma_{\mathrm{CR}}^2\approx\frac{1}{N_{cT}^2}N_{cT}
\frac{kT}{C_{\mathrm{tot}}/N_{cT}}=\frac{kT}{C_{\mathrm{tot}}}.
\label{eq:ktc}
\end{equation}

Thus, under ideal settling and equal-weight averaging, the first-order
sampling noise can remain comparable to that of a conventional sampling
event using the full capacitance \(\left( C_{tot} \right)\). This is an
important circuit-level advantage of charge-domain temporal compression.
In practice, the realized noise and distortion also depend on capacitor
matching, switch resistance, settling time, charge injection, clock
feedthrough, and residue-generation accuracy.

Because the ADC output represents a coded analog measurement, circuit
nonidealities affect the effective sensing matrix used during RF
recovery. Relevant nonidealities include capacitor mismatch, incomplete
settling of selected CDAC subsets, PRBS switch timing skew, charge
injection, clock feedthrough, leakage between sampling and charge
redistribution, comparator kickback, residue-amplifier gain error, and
common-mode shifts after charge redistribution.

The reconstruction algorithm does not require the ideal mathematical
matrix alone; it requires the effective mapping between the original RF
samples and the measured ADC codes. Static gain terms can be absorbed
into the effective matrix or corrected by digital calibration. More
complex errors, such as PRBS-dependent mismatch, incomplete charge
redistribution, or sample-dependent charge injection, can produce
structured reconstruction artifacts unless they are minimized by circuit
design or included in calibration.

In the present prototype, the charge-redistribution control phase is a
major performance-limiting factor. The charge-redistribution pulse is
generated using delay-cell-based timing, and its pulse width and drive
strength are sensitive to parasitic loading from switches, routing, and
CDAC nodes. Residual voltage error after the charge-redistribution phase
is amplified by the inter-stage residue amplifier and appears as
distortion or noise at the second-stage input. This mechanism is
consistent with the measured SNDR being below the original design
target.

In future designs, a revised implementation should generate the
charge-redistribution phase from a more controlled clocking edge or from
additional phases of the non-overlapping clock generator. Stronger
buffering, better-defined pulse width, and layout-aware timing closure
would reduce residual charge-redistribution error. This improvement is
particularly important in compression modes because the compressed
measurement depends directly on accurate charge-domain summation before
quantization.

The present study has several limitations. First, the prototype
demonstrates temporal compression only. It does not reduce
receive-channel count or analog front-end count. Second, RF recovery is
performed off chip, so the computational cost and memory requirements of
real-time reconstruction remain to be optimized. Third, the current
validation uses FIELD-II-generated RF data applied to the ADC as
analog-equivalent input waveforms rather than a fully integrated
probe-to-ADC ultrasound receive chain. Fourth, the regularization
parameters were selected empirically; future work should evaluate
noise-, signal-energy-, and acquisition-dependent parameter selection for
robust operation across probes, depths, steering angles, and tissue
types. Finally, the cyst phantom provides an important
speckle-rich validation, but broader testing across additional phantoms,
in vivo datasets, and different imaging depths would be required to
establish clinical-grade performance.

Although the present paper focuses on temporal CS acquisition, the same
mathematical framework can be extended to space-time compression. In a
full space-time architecture, the measurement matrix would mix samples
across both time and receive-channel dimensions. Such an approach could
reduce ADC output data rate, receive-channel count, cable count, and
analog front-end footprint simultaneously. A generic space-time model
can be written as

\begin{equation}
\mathbf{y}=\mathbf{C}_{ST}\mathbf{x}.
\label{eq:spacetime}
\end{equation}

where $\mathbf{C}_{ST}$ is no longer block diagonal with respect
to receive channel index. Instead, it may encode spatial mixing,
temporal mixing, or both. This is a natural direction for future
integrated ultrasound receivers, especially for catheter-based,
wearable, and high-channel-count 3-D imaging systems.
\bibliographystyle{IEEEtran}
\bibliography{references}

@article{chen2021integrated,
  author  = {C. Chen and M. A. P. Pertijs},
  title   = {Integrated Transceivers for Emerging Medical Ultrasound Imaging Devices: A Review},
  journal = {IEEE Open Journal of the Solid-State Circuits Society},
  volume  = {1},
  pages   = {104--114},
  year    = {2021}
}

@article{giangrossi2022requirements,
  author  = {C. Giangrossi and A. Ramalli and A. Dallai and D. Mazierli and V. Meacci and E. Boni and P. Tortoli},
  title   = {Requirements and Hardware Limitations of High-Frame-Rate 3-D Ultrasound Imaging Systems},
  journal = {Applied Sciences},
  volume  = {12},
  number  = {13},
  pages   = {6562},
  year    = {2022},
  doi     = {10.3390/app12136562}
}

@article{black1994cmos,
  author  = {W. C. Black and D. N. Stephens},
  title   = {{CMOS} Chip for Invasive Ultrasound Imaging},
  journal = {IEEE Journal of Solid-State Circuits},
  volume  = {29},
  number  = {11},
  pages   = {1381--1387},
  year    = {1994}
}

@article{gurun2014singlechip,
  author  = {G. Gurun and C. Tekes and J. Zahorian and M. Karaman and P. E. Hasler and F. L. Degertekin},
  title   = {Single-Chip {CMUT}-on-{CMOS} Front-End System for Real-Time Volumetric {IVUS} and {ICE} Imaging},
  journal = {IEEE Transactions on Ultrasonics, Ferroelectrics, and Frequency Control},
  volume  = {61},
  number  = {2},
  pages   = {239--250},
  year    = {2014}
}

@article{chen2017frontend,
  author  = {C. Chen and others},
  title   = {A Front-End {ASIC} With Receive Sub-Array Beamforming Integrated With a 32 $\times$ 32 {PZT} Matrix Transducer for 3-D Transesophageal Echocardiography},
  journal = {IEEE Journal of Solid-State Circuits},
  volume  = {52},
  number  = {4},
  pages   = {994--1006},
  year    = {2017}
}

@article{rezvanitabar2022integrated,
  author  = {A. Rezvanitabar and C. Tekes and F. L. Degertekin and M. Ghovanloo},
  title   = {Integrated Hybrid Sub-Aperture Beamforming and Time-Division Multiplexing for Massive Readout in Ultrasound Imaging},
  journal = {IEEE Transactions on Biomedical Circuits and Systems},
  volume  = {16},
  number  = {5},
  pages   = {972--980},
  year    = {2022}
}

@article{montaldo2009coherent,
  author  = {G. Montaldo and M. Tanter and J. Bercoff and N. Benech and M. Fink},
  title   = {Coherent Plane-Wave Compounding for Very High Frame Rate Ultrasonography and Transient Elastography},
  journal = {IEEE Transactions on Ultrasonics, Ferroelectrics, and Frequency Control},
  volume  = {56},
  number  = {3},
  pages   = {489--506},
  year    = {2009}
}

@article{candes2008introduction,
  author  = {E. J. Cand\`es and M. B. Wakin},
  title   = {An Introduction to Compressive Sampling},
  journal = {IEEE Signal Processing Magazine},
  volume  = {25},
  number  = {2},
  pages   = {21--30},
  year    = {2008}
}

@article{candes2006robust,
  author  = {E. J. Cand\`es and J. Romberg and T. Tao},
  title   = {Robust Uncertainty Principles: Exact Signal Reconstruction From Highly Incomplete Frequency Information},
  journal = {IEEE Transactions on Information Theory},
  volume  = {52},
  number  = {2},
  pages   = {489--509},
  year    = {2006}
}

@article{donoho2006compressed,
  author  = {D. L. Donoho},
  title   = {Compressed Sensing},
  journal = {IEEE Transactions on Information Theory},
  volume  = {52},
  number  = {4},
  pages   = {1289--1306},
  year    = {2006}
}

@article{tur2011innovation,
  author  = {R. Tur and Y. C. Eldar and Z. Friedman},
  title   = {Innovation Rate Sampling of Pulse Streams With Application to Ultrasound Imaging},
  journal = {IEEE Transactions on Signal Processing},
  volume  = {59},
  number  = {4},
  pages   = {1827--1842},
  year    = {2011}
}

@article{wagner2012compressed,
  author  = {N. Wagner and Y. C. Eldar and Z. Friedman},
  title   = {Compressed Beamforming in Ultrasound Imaging},
  journal = {IEEE Transactions on Signal Processing},
  volume  = {60},
  number  = {9},
  pages   = {4643--4657},
  year    = {2012}
}

@article{chernyakova2014fourier,
  author  = {T. Chernyakova and Y. C. Eldar},
  title   = {Fourier-Domain Beamforming: The Path to Compressed Ultrasound Imaging},
  journal = {IEEE Transactions on Ultrasonics, Ferroelectrics, and Frequency Control},
  volume  = {61},
  number  = {8},
  pages   = {1252--1267},
  year    = {2014},
  doi     = {10.1109/TUFFC.2014.3032}
}

@article{liebgott2013prebeamformed,
  author  = {H. Liebgott and R. Prost and D. Friboulet},
  title   = {Pre-Beamformed {RF} Signal Reconstruction in Medical Ultrasound Using Compressive Sensing},
  journal = {Ultrasonics},
  volume  = {53},
  number  = {2},
  pages   = {525--533},
  year    = {2013}
}

@article{lorintiu2015dictionary,
  author  = {O. Lorintiu and H. Liebgott and M. Alessandrini and O. Bernard and D. Friboulet},
  title   = {Compressed Sensing Reconstruction of 3-D Ultrasound Data Using Dictionary Learning and Line-Wise Subsampling},
  journal = {IEEE Transactions on Medical Imaging},
  volume  = {34},
  number  = {12},
  pages   = {2467--2477},
  year    = {2015}
}

@inproceedings{besson2017approach,
  author    = {A. Besson and R. E. Carrillo and D. Perdios and M. Arditi and Y. Wiaux and J.-P. Thiran},
  title     = {A Compressed-Sensing Approach for Ultrasound Imaging},
  booktitle = {Proceedings of the Signal Processing With Adaptive Sparse Structured Representations (SPARS) Workshop},
  address   = {Lisbon, Portugal},
  month     = {June},
  year      = {2017}
}

@article{mitrovic2020hardware,
  author  = {J. Mitrovic and Z. Ignjatovic and L. L. Pietra and W. J. Sehnert and V. Dogra},
  title   = {Compressed Sensing for Reduced Hardware Footprint in Medical Ultrasound},
  journal = {Ultrasonics},
  volume  = {108},
  pages   = {106214},
  year    = {2020},
  doi     = {10.1016/j.ultras.2020.106214}
}

@article{anand2021practical,
  author  = {R. Anand and A. K. Thittai},
  title   = {Towards Practical Implementation of the Compressed Sensing Framework for Multi-Element Synthetic Transmit Aperture Imaging},
  journal = {Ultrasonics},
  volume  = {112},
  pages   = {106354},
  year    = {2021},
  doi     = {10.1016/j.ultras.2021.106354}
}

@article{mamistvalov2022convolutional,
  author  = {A. Mamistvalov and Y. C. Eldar},
  title   = {Compressed Fourier-Domain Convolutional Beamforming for Sub-Nyquist Ultrasound Imaging},
  journal = {IEEE Transactions on Ultrasonics, Ferroelectrics, and Frequency Control},
  volume  = {69},
  number  = {2},
  pages   = {489--499},
  year    = {2022},
  doi     = {10.1109/TUFFC.2021.3123079}
}

@article{liao2025mvm,
  author  = {H.-C. Liao and others},
  title   = {Compressive Sensing Photoacoustic Imaging Receiver With Matrix-Vector-Multiplication {SAR ADC}},
  journal = {IEEE Journal of Solid-State Circuits},
  volume  = {60},
  number  = {11},
  pages   = {3895--3907},
  year    = {2025},
  doi     = {10.1109/JSSC.2025.3603157}
}

@article{matrone2015dmas,
  author  = {G. Matrone and A. S. Savoia and G. Caliano and G. Magenes},
  title   = {The Delay Multiply and Sum Beamforming Algorithm in Ultrasound B-Mode Medical Imaging},
  journal = {IEEE Transactions on Medical Imaging},
  volume  = {34},
  number  = {4},
  pages   = {940--949},
  year    = {2015},
  doi     = {10.1109/TMI.2014.2371235}
}

@inproceedings{matrone2016planewave,
  author    = {G. Matrone and A. S. Savoia and G. Caliano and G. Magenes},
  title     = {Ultrasound Plane-Wave Imaging With Delay Multiply and Sum Beamforming and Coherent Compounding},
  booktitle = {2016 38th Annual International Conference of the IEEE Engineering in Medicine and Biology Society (EMBC)},
  pages     = {3223--3226},
  year      = {2016},
  doi       = {10.1109/EMBC.2016.7591415}
}

@article{jensen1992calculation,
  author  = {J. A. Jensen and N. B. Svendsen},
  title   = {Calculation of Pressure Fields From Arbitrarily Shaped, Apodized, and Excited Ultrasound Transducers},
  journal = {IEEE Transactions on Ultrasonics, Ferroelectrics, and Frequency Control},
  volume  = {39},
  number  = {2},
  pages   = {262--267},
  year    = {1992}
}

@inproceedings{jensen1996field,
  author    = {J. A. Jensen},
  title     = {{FIELD}: A Program for Simulating Ultrasound Systems},
  booktitle = {Proceedings of the 10th Nordic-Baltic Conference on Biomedical Imaging},
  journal   = {Medical \& Biological Engineering \& Computing},
  volume    = {34},
  number    = {Supplement 1, Part 1},
  pages     = {351--353},
  year      = {1996}
}

\end{document}